\documentclass[a4paper,11pt]{article}
\pdfoutput=1 

\usepackage{jcappub} 

\usepackage{orcidlink}

\usepackage[T1]{fontenc} 
\usepackage{multirow}
\usepackage{makecell}
\usepackage{subcaption}
\usepackage[normalem]{ulem}
\usepackage{graphicx}

\graphicspath{{./figures/}}

\newcommand{\vk}{{\mathbf k}}
\newcommand{\vx}{{\mathbf x}}

\def \FFdual {F_{\mu\nu}\tilde{F}^{\mu\nu}}

\newcommand{\CLns}{{\tt ${\mathcal C}$osmo${\mathcal L}$attice}}

\title{\boldmath Comparative study of the strong backreaction regime in axion inflation: the effect of the potential}

\author[a,b]{Joanes Lizarraga\orcidlink{0000-0002-1198-3191},}
\author[a,b]{Carmelo López-Mediavilla\orcidlink{0009-0007-8327-2472}}
\author[a,b]{and Ander Urio\orcidlink{0000-0002-0238-8390}}

\affiliation[a]{Department of Physics, University of Basque Country, UPV/EHU, 48080, Bilbao, Spain}
\affiliation[b]{EHU Quantum Center, University of the Basque Country UPV/EHU, Leioa, 48940 Biscay, Spain}

\emailAdd{joanes.lizarraga@ehu.eus}
\emailAdd{carmelo.lopez@ehu.eus}
\emailAdd{ander.urio@ehu.eus}

\abstract{Recent works have demonstrated the necessity of capturing the local inhomogeneous physics in axion inflation, and showed new genuine features, most notably the extension of the inflationary period dictated by an electromagnetic slow-roll phase. In this work, we further investigate the model by performing a systematic study of the effect of the inflationary potential in the dynamics during the strong backreaction regime. The results indicate that the novel features associated with the local backreaction are universal and intrinsic to the model, hence independent on the choice of inflationary potential. We find that the main quantitative differences between the different choices manifest in the lengthening of inflation. We discuss the possible observational impact of this. Finally, we assess the possible reconciliation of the homogeneous backreaction method with fully inhomogeneous lattice techniques, and obtain that the former fails to provide a correct description for the regime studied in this work.}

\begin{document}
\maketitle
\flushbottom

\section{Introduction}

Axion inflation, particularly in the form of Abelian axion inflation, has emerged as a notably successful framework for embedding inflationary dynamics within a more robust theoretical context. While single-field inflation remains a minimal and observationally consistent approach \cite{Martin:2013tda,Planck:2018jri}, challenges in UV completion persist \cite{Lyth:1998xn,Baumann:2009ds,Pajer:2013fsa,Baumann:2014nda}. The identification of axion-like particles (ALPs), protected by shift symmetries, as the inflaton offers a compelling solution by naturally preserving the flatness of the inflaton potential \cite{Freese:1990rb,Adams:1992bn}. In the Abelian axion inflation model \cite{Anber:2006xt,Anber:2009ua}, the axion couples to a U(1) gauge field via a Chern-Simons interaction, enabling efficient (tachyonic) energy transfer and triggering the so-called chiral instabilities in the gauge field \cite{Turner:1987vd,Garretson:1992vt,Anber:2006xt,Anber:2009ua,Barnaby:2010vf,Adshead:2013qp,Cheng:2015oqa}. This leads to a rich phenomenology, including, for instance, non-Gaussian curvature perturbations \cite{Barnaby:2010vf,Barnaby:2011vw,Cook:2011hg, Barnaby:2011qe,Pajer:2013fsa}, chiral gravitational waves \cite{Sorbo:2011rz, Barnaby:2011qe, Cook:2013xea,Adshead:2013qp,Bastero-Gil:2022fme,Garcia-Bellido:2023ser}, and efficient (p)reheating \cite{Adshead:2015pva,Cuissa:2018oiw,Adshead:2023mvt} with sizeable post-inflationary GW production~\cite{Adshead:2018doq,Adshead:2019igv,Adshead:2019lbr}.

The dynamics of the model is highly non-linear and has been approached using various levels of approximation. A central aspect is the electromagnetic backreaction, which is intrinsically inhomogeneous. Despite this, the standard approach has been to assume a homogeneous backreaction  \cite{Cheng:2015oqa,Notari:2016npn,DallAgata:2019yrr,Sobol:2019xls,Domcke:2020zez,Gorbar:2021rlt,Peloso:2022ovc,Durrer:2023rhc,vonEckardstein:2023gwk,Galanti:2024jhw,Durrer:2024ibi,vonEckardstein:2024tix} and, in fact, most of the phenomenological predictions in the literature to date rely on this homogeneous approximation. Yet, this homogenization is not grounded in the physical characteristics of the model, but rather in the goal of improving its numerical solvability. 

In recent years lattice techniques have become increasingly popular as a way to capture the local backreaction and resolve the genuine non-linear dynamics of the model, both in preheating scenarios \cite{Adshead:2015pva,Adshead:2018doq,Adshead:2019igv,Adshead:2019lbr,Cuissa:2018oiw} and during inflation itself \cite{Caravano:2021bfn,Caravano:2022epk,Figueroa:2023oxc,Figueroa:2024rkr,Sharma:2024nfu}. The latter includes the works \cite{Figueroa:2023oxc,Figueroa:2024rkr}, which were authored by some of  the present authors. These works have represented a paradigm shift in the study of axion inflation in the strong backreaction regime, as they point out the shortcomings of the homogeneous approximation and uncover novel features of the dynamics, including a significant extension of the inflationary period driven by a late magnetic dominance phase named as electromagnetic slow-roll, the formation of inflaton inhomogeneities that affect the UV sensitivity of the model, and a reduction of the chiral imbalance and loss of the transversality of the gauge field.

Most of the axion inflation literature has been built upon the analysis of the evolution under chaotic type potentials, or in the context of the original potential from natural inflation. These realizations, however, are either excluded or disfavoured by current CMB constraints \cite{Planck:2018jri,BICEP:2021xfz}. This work aims at complementing and extending our previous works \cite{Figueroa:2023oxc,Figueroa:2024rkr} by considering a broad set of single-field inflationary potentials so as to analyse their effect in the dynamics. An analysis along these lines was performed in \cite{Adshead:2019igv,Adshead:2019lbr}, where the impact of the potential was studied with a focus on the efficiency of preheating and subsequent gravitational wave generation. In contrast, this work focuses on the inflationary dynamics in the strong backreaction regime, as in \cite{Figueroa:2023oxc,Figueroa:2024rkr}, considering larger values of the coupling constants.

A prime objective of this study is to assess the universality of the novel features of the strong backreaction regime of axion inflation. To this end we perform a systematic study of the dynamics into such regime for the whole set of different potentials. In addition, this work also tests the range of validity of homogeneous backreaction procedures. We perform the corresponding comparison in a variety of setups, \textit{i.e.} different potentials and coupling strengths, in order to challenge whether there is a situation where both techniques tend to reconcile.

The paper is organized as follows: after a short introduction to the model in Sec.~\ref{sec:AxionInflation}, we summarize the discretisation scheme used to solve the system, along with the definition of appropriate initial conditions in Sec.~\ref{subsec:latticeForm}. In Sec.~\ref{subsec:potentials}, we motivate the need to go beyond the quadratic case and introduce the selected potentials for our study. For each potential, we present the relevant parameters and identify the relevant ranges in Sec.~\ref{subsec:ICandRunParams}. The results are then presented from two complementary perspectives in Sec.~\ref{sec:results}. First, in Sec.~\ref{subsec:results_dynamics}, we analyse the behaviour of all potentials under the local backreaction approach, allowing us to uncover differences in their dynamics and to classify them into similarity groups. Second, in Sec.~\ref{subsec:results_homo}, we compare the dynamics of the system for each potential under both the local backreaction and the homogeneous backreaction approaches.

We note that throughout this work we use the metric signature $(-,+,+,+)$, and define the reduced Planck mass as $m_{\rm{p}} = 2.435 \cdot 10^{18}~\rm{GeV}$.


\section{Model and numerical implementation}
\label{sec:AxionInflation}

We consider the minimal axion inflation model, where a shift-symmetric pseudoscalar field $\phi$ plays the role of the inflaton and is coupled to a U(1) gauge sector $A_\mu$ through a Chern-Simons term \cite{Anber:2006xt,Anber:2009ua,Barnaby:2010vf,Barnaby:2011qe,Sorbo:2011rz,Barnaby:2011vw,Meerburg:2012id,Linde:2012bt,Cook:2013xea,Cheng:2015oqa,Anber:2015yca,Notari:2016npn,Domcke:2018eki,DallAgata:2019yrr,Sobol:2019xls,Domcke:2019qmm,Domcke:2020zez,Gorbar:2021rlt,Peloso:2022ovc,Cado:2022pxk,Bastero-Gil:2022fme,Durrer:2023rhc,vonEckardstein:2023gwk,Garcia-Bellido:2023ser,Durrer:2024ibi,Galanti:2024jhw,vonEckardstein:2024tix,Adshead:2015pva,Adshead:2018doq,Adshead:2019igv,Adshead:2019lbr,Cuissa:2018oiw,Domcke:2023tnn,Adshead:2023mvt,Fujita:2025zoa,He:2025ieo,Kume:2025lvz,Sharma:2024nfu,Corba:2025reo,Ozsoy:2024apn,Corba:2024tfz,Gorbar:2023zla,Unal:2023srk,Caravano:2022epk,Caravano:2021bfn,Figueroa:2024rkr,Figueroa:2023oxc}:

\begin{equation}\label{eq:actionCONT}
    S= \int {\rm d}x^4 \sqrt{-g}\left[\frac{1}{2}m^2_{\rm p}R-\frac{1}{2}\partial_\mu \phi\partial^\mu\phi-V(\phi) - \frac{1}{4}F_{\mu\nu}F^{\mu\nu} + \frac{\alpha_{\Lambda}}{4}\frac{\phi}{m_{\rm{p}}} \FFdual \right]\;.
\end{equation}
The matter sector is minimally coupled to gravity, and the interaction therein between the scalar and gauge sectors is ruled by the dimensionless coupling $\alpha_{\Lambda}\equiv m_{\rm p}/\Lambda$, where $\Lambda$ is the axion decay constant. The field strength and its dual counterpart of the gauge field $A_{\mu}$ are the standard
\begin{equation}\label{eq:FieldStrAndDual}
    F_{\mu\nu}\equiv\partial_\mu A_\nu-\partial_\nu A_\mu\,,~~~
\tilde F_{\mu \nu}\equiv \frac{1}{2}\epsilon_{\mu\nu\rho\sigma}F^{\rho\sigma}\,,
\end{equation}
where $\epsilon_{\mu\nu\rho\sigma}$ is the fully antisymmetric Levi-Civita tensor in curved spacetimes, such that $\epsilon_{0123} = 1 / \sqrt{-g}$.

The shift symmetry is typically explicitly broken by the potential $V(\phi)$, which is responsible for driving the exponential expansion of the universe during inflation. A UV complete description of the model should include the fundamental process behind its emergence. The main objective of this work is just to study the effect of different shapes into the dynamics during the strong backreaction. Therefore, we keep a completely agnostic perspective about the origin of the concrete form of the possible potentials and assume that they are set by some external processes. We present in Sec.~\ref{subsec:potentials} the set of potentials studied in this work.

In an unperturbed spatially flat Friedmann-Lemaître-Robertson-Walker (FLRW) background, with homogeneous scale factor $a$, the equations of motion (EOM) for the matter sector in the temporal gauge ($A_0=0$) read as
\begin{align}
	\ddot{\phi} &=  -3H\dot{\phi} + \frac{1}{a^2} \vec{\nabla}^2 \phi - \frac{dV(\phi)}{d\phi} + \frac{\alpha_\Lambda}{a^3 m_{\rm{p}}} \vec{E}\cdot\vec{B} \label{eq:EOMscalarCONT} \; , \\[4pt] 
	\dot{\vec{E}} &= - H \vec{E} - \frac{1}{a^2} \vec{\nabla} \times \vec{B} - \frac{\alpha_\Lambda}{am_{\rm{p}}} \left( \dot{\phi} \vec{B} - \vec{\nabla} \phi \times \vec{E} \right) \label{eq:EOMgaugeCONT} \; , 
  \\[4pt]  
    \vec{\nabla} \cdot \vec{E} &= - \frac{\alpha_\Lambda}{a m_{\rm{p}}} \vec{\nabla} \phi \cdot \vec{B} \; , \label{eq:gaussCONT}
\end{align}
where $\dot{}=\frac{d}{dt}$ represents differentiation with respect cosmic time and $H(t)=\dot{a}/a$ is the Hubble parameter. We use the vectorial form of the EOM, for which we have defined the electric and magnetic field as
\begin{equation}\label{eq:ElecMagnetDefCONT}
    E_i\equiv\dot A_i\,,\quad B_i\equiv\epsilon_{ijk}\partial_j A_k\,.
\end{equation}

The spacetime background dynamics is set by the Friedmann equations
\begin{align}
\ddot{a}&=-\frac{a}{3m_{\rm{p}}^2}\big( 2\rho_{\rm K}-\rho_{\rm V}+\rho_{\rm EM} \big)\,,
\label{eq:friedman1CONT}
\\[4pt] 
H^2&=\frac{1}{3m_{\rm{p}}^2}\big(\rho_{\rm K}+\rho_{\rm G}+\rho_{\rm V}+\rho_{\rm EM}\big)\,,
\label{eq:friedman2CONT}
\end{align}
where different homogeneous contributions to the energy density from the inflaton and the gauge field are
\begin{equation}
\begin{gathered}\label{eq:enDensCONT}
\rho_{\rm K} \equiv \frac{1}{2}\langle\dot{\phi}^2\rangle\; , \quad \rho_{\rm G} \equiv \frac{1}{2a^2}\langle(\vec\nabla\phi)^2\rangle\; ,
\\[4pt]
\rho_{\rm V} \equiv \langle V\rangle\;, \quad\rho_{\rm EM} \equiv \frac{1}{2a^4}\langle a^2\vec E^2+\vec B^2\rangle \;,
\end{gathered}
\end{equation}
with $\langle ... \rangle$ denoting the volume average, and the subscripts $\rm{K}$, $\rm{V}$, and $\rm{G}$ referring to the kinetic, potential, and gradient energy densities of the inflation, respectively, and $\rm{EM}$ to the electromagnetic energy density.

The Eqs.~(\ref{eq:EOMscalarCONT}), (\ref{eq:EOMgaugeCONT}) and (\ref{eq:friedman1CONT}) form the dynamical set of EOM, whereas Eqs.~(\ref{eq:gaussCONT}) and (\ref{eq:friedman2CONT}) represent the Gauss's constraint and the Hubble's constraint, respectively.

This coupled and highly non-linear set of equations have been studied under different levels of approximation, which are valid within certain regimes of the evolution. Here we just present a brief review on them and refer the reader to our previous works \cite{Figueroa:2023oxc,Figueroa:2024rkr} for a throughout discussion and detailed comparisons.

The first layer of complexity is set by studying the model under the \textit{backreaction-less} solution, where the inflaton and the spacetime background evolve without being affected by the presence of the gauge fields, hence the excitations of the gauge field are sourced passively. A characteristic feature of this approximation is that only one of the possible transverse polarisations of the gauge field is excited and, indeed, acquires a negative effective mass inducing a tachyonic instability \cite{Turner:1987vd,Garretson:1992vt,Anber:2006xt,Anber:2009ua}. This instability is triggered by the inflaton's momentum, and it is called the chiral instability of axion inflation. In this regime the solution admits an analytical form 
\begin{equation}
A^+(t,{\bf k}) \simeq \frac{1}{\sqrt{2k}}\left(\frac{k}{2\xi aH}\right)^{1/4} e^{\pi\xi - 2\sqrt{2\xi k/(aH)}} \quad \text{with} \ \xi \equiv \frac{|\langle\dot{\phi}\rangle|}{2 H \Lambda }\,.
\label{eqn:amplifiedAplus}
\end{equation}
The subscript $^+$ refers to a component of the chiral decomposition of the gauge field,

\begin{equation}\label{eq:gaugeFourierChiral}
    \vec A(t,\vx) = \sum_{\lambda=\pm} \int \frac{{\rm d}^3k}{(2\pi)^3}  A^\lambda(t,{\bf k})\vec{\varepsilon}^{\,\lambda}(\hat\vk)  e^{i\vk\cdot\vx}\, ,
\end{equation}
where the vectors $\vec{\varepsilon}^{\,\lambda}$, with $\lambda=\pm$, form the chiral basis and satisfy
\begin{equation}
\begin{gathered}
  \hat \vk\cdot\vec\varepsilon^{\,\lambda}(\hat\vk)=0\,,\quad
\hat \vk \times \vec \varepsilon^{\,\lambda}(\hat\vk)=-i\lambda \vec\varepsilon^{\,\lambda}(\hat\vk)\,,\\[4pt]
\varepsilon_i^{\,\lambda}(\hat\vk)^*=\varepsilon_i^{\,\lambda}(-\hat\vk)\,,\quad \vec \varepsilon^{\,\lambda'}\hspace*{-1mm}(\hat\vk)\cdot\vec\varepsilon^{\,\lambda}(\hat\vk)^* = \delta_{\lambda\lambda'}\, .
\end{gathered}
\end{equation}

They can be explicitly defined with respect a reference direction, $\hat{e}_3$ of the cartesian basis
\begin{equation}
    \varepsilon_i^\pm(\theta,\varphi) \equiv (u_i(\theta,\varphi) \pm iv_i(\theta,\varphi))/\sqrt{2}\, ,
    \label{eq:chiral_vec}
\end{equation}
with
\begin{align}
    \vec v(\hat\vk) &\equiv (\hat e_3 \times \hat {\bf k})/|\hat e_3 \times \hat {\bf k}| = (-\sin\varphi,\cos\varphi,0)\;, \\[4pt]
    \vec u(\hat\vk) &\equiv  \vec v \times \hat {\bf k} = (\cos\theta\cos\varphi,\cos\theta\sin\varphi,-\sin\theta)\;.
\end{align}
where $\hat{\textbf{k}}=(\sin\theta\cos\varphi,\sin\theta\sin\varphi,\cos\theta)$.

The next level of approximation considers the backreaction of the gauge fields into the scalar sector dynamics and their effect in the expansion of the universe. Aside lattice techniques, which we will describe in the next subsection, this has been accomplished by considering expectation values of $\vec{E} \cdot \vec{B} \rightarrow \langle \vec{E} \cdot \vec{B} \rangle$ in Eq.~(\ref{eq:EOMscalarCONT}). This \textit{homogeneous backreaction} approach, thereby includes the backreaction while maintaining $\phi$ homogeneous, \textit{i.e.} $\vec{\nabla}\phi~=~0$. Two are the main methodologies to accomplish this, the integro-differential approach \cite{Cheng:2015oqa,Notari:2016npn,DallAgata:2019yrr,Domcke:2020zez,Bastero-Gil:2022fme} and the Gradient Expansion Formalism \cite{Sobol:2019xls,Gorbar:2021rlt,Durrer:2023rhc,vonEckardstein:2023gwk,Durrer:2024ibi,vonEckardstein:2024tix,Garcia-Bellido:2023ser}. While the first method solves an integro-differential set of equations for the scalar and gauge sector mode-by-mode, the latter considers the evolution of suitably constructed bilinear functions of the electromagnetic components. Both approaches show equivalent results, and can be summarized by the emergence of a resonant amplification of the excited chiral mode, which manifests as an oscillatory behaviour in the inflaton velocity \cite{Domcke:2020zez} and the gauge field spectrum \cite{Figueroa:2023oxc, Figueroa:2024rkr}. This can be understood as a consequence of the time delay between the peak of the excitation rate of $A^+$, which lays at slightly sub-Hubble scales and the dominant part of $\langle \vec{E} \cdot \vec{B} \rangle$ located at slightly super-Hubble modes \cite{Domcke:2020zez}. Even though the dynamics is substantially different from that of the backreaction-less case, the theory remains completely chiral.

In our previous works \cite{Figueroa:2023oxc,Figueroa:2024rkr} we compared the validity of the different approximations against the description given by fully inhomogeneous lattice simulations, and demonstrated that they fail to provide an adequate description of the system's dynamics when a quadratic (chaotic) potential is considered. In particular, solving for the full inhomogeneous system showed that inflationary periods with significantly different extra number of e-folds emerge, with a partial scale dependent chirality balance restoration.

\subsection{Discretisation procedure}
\label{subsec:latticeForm}

In order to account for the backreaction with all genuine inhomogeneities, lattice techniques must be employed\footnote{See \cite{Domcke:2023tnn} for a proposal that includes inhomogeneities perturbatively.}. We follow the formalism proposed in \cite{Figueroa:2023oxc,Figueroa:2024rkr}, which is itself an adaptation of \cite{Figueroa:2017qmv,Cuissa:2018oiw}, for an exactly shift-symmetric and gauge-invariant procedure. We refer the reader to those references for detailed descriptions, validations and discussions on this lattice prescription. In this section, we limit ourselves to briefly summarize the definitions and properties of the formalism, and present the discretised equations of motion.

We discretise the model action for the matter sector by the following hybrid continuum-discrete lattice action,

\begin{multline}\label{eq:latticeAction}
	S_{\rm m}^{\rm L} = \int dt \sum_{\textbf{n}}\delta x^3 \left\{ \frac{1}{2}a^3\pi^2_{\phi} - \sum_i a\frac{1}{2}(\Delta^+_i\phi)^2 -a^3V(\phi) + \right. \\
	\left. + \sum_i \frac{1}{2}a\left(E_i^2 - a^{-2}B_i^2\right)  + \frac{\alpha_\Lambda\phi}{m_{\rm{p}}} \sum_i E_i^{(2)}B_i^{(4)} \right\} \; ,
\end{multline}
where we consider the scalar field $\phi(\textbf{n})$ to live on the lattice \textit{sites} $\textbf{n}$, while the gauge field $A_i(\textbf{n}+\hat{\imath}/2)$ ($A_0 = 0$ in the temporal gauge) resides on the \textit{links}. The discrete spatial derivatives are computed through the usual forward or backward lattice derivatives: $\Delta^{\pm}_i\phi(\textbf{n}+\hat{\imath}/2)=\frac{1}{dx}[\pm\phi(\textbf{n}\pm\hat{\imath})\mp\phi(\textbf{n})]$. 

In this formalism we use two different definitions for the electric and magnetic fields:
\begin{equation}
\begin{gathered}
    E_i(\textbf{n}+\hat{\imath}/2) \equiv 
\dot A_i\,, \quad ; \quad E_{i}^{(2)} (\textbf{n}) \equiv  \frac{1}{2}\left(E_{i}+E_{i,-i}\right)\,,  \\[4pt]
B_i(\textbf{n}+\hat{\jmath}/2+\hat{k}/2) \equiv \sum_{j,k}\epsilon_{ijk}\Delta^+_jA_k\,, \quad \  ; \quad  B_{i}^{(4)} (\textbf{n})\equiv \frac{1}{4}\left(B_{i}+B_{i,-j}+B_{i,-k}+B_{i,-j-k}\right)\, ,
\end{gathered}
\label{eq:latticeStandEandB}
\end{equation}
with subscript $(...)_{(...),i}$ indicating displacement of a unit in the $i$th direction. The definitions on the left $\{E_i,B_i\}$ are the standard ones, whereas $\{E^{(2)}_i,B^{(4)}_i\}$ on the right consider \textit{improved} versions, \textit{i.e.} spatially symmetrized averages. The latter redefinitions are necessary to achieve exact gauge invariance and shift symmetry on the lattice as demonstrated in \cite{Figueroa:2017qmv}.

For numerical convenience we work in dimensionless variables that set the program variables of the system:
\begin{equation}    d\tilde{x}^{\mu}=\omega_{*}dx^{\mu}\;,\quad \tilde{\phi}=\frac{\phi}{f_{*}}\;,\quad \tilde{A}_{i}=\frac{A_i}{\omega_{*}}\;,
\end{equation}
where $f_{*}$ and $\omega_{*}$ set the typical energy scale and timescale of the model. Their choice is, in principle, arbitrary, but some values can be used to optimize the numerical treatment. In this work, where we make a systematic comparison for different inflationary potentials, different scales have been considered. We summarize them in Table~\ref{tab:potentials}.

The discretised field EOM are obtained by variation of the action (\ref{eq:latticeAction}), which in dimensionless variables

\begin{align}
	\dot{\tilde{\pi}}_{\phi} &=  -3\tilde{H}\tilde{\pi}_{\phi} + \frac{1}{a^2} \sum_i \tilde{\Delta}_i^-\tilde{\Delta}_i^+ \tilde{\phi} - \frac{d\tilde{V}(\tilde{\phi})}{d\tilde{\phi}} + \left[\frac{\omega^2_{*}}{f_{*}m_{\rm{p}}}\right]\frac{\alpha_\Lambda}{a^3} \sum_i \tilde{E}_i^{(2)}\tilde{B}_i^{(4)} \; , \label{eq:EOMscalarLATprog}\\[4pt]
	 \dot{\tilde{E}}_i &=  \tilde{H}\tilde{E}_i - \frac{1}{a^2} \sum_{j,k} \epsilon_{ijk} \tilde{\Delta}_j^- \tilde{B}_k - \left[\frac{f_{*}}{m_{\rm{p}}}\right] \frac{\alpha_\Lambda}{2a}\left(\tilde{\pi}_{\phi} \tilde{B}_i^{(4)} + \tilde{\pi}_{\phi,+i}\tilde{B}^{(4)}_{i,+i} \right) \nonumber \\[2pt]
	& + \left[\frac{f_{*}}{m_{\rm{p}}}\right]\frac{\alpha_\Lambda}{4a} \sum_\pm \sum_{j,k} \epsilon_{ijk}  \left\{ \left[ (\tilde{\Delta}_j^\pm \tilde{\phi}) \tilde{E}_{k,\pm j}^{(2)} \right]_{+i} +  \left[ (\tilde{\Delta}_j^\pm \tilde{\phi}) \tilde{E}_{k,\pm j}^{(2)}  \right]   \right\} \; , \label{eq:EOMgaugeLATprog}\\[4pt]
    	\sum_i \tilde{\Delta}_i^- \tilde{E}_i &= - \left[\frac{f_{*}}{m_{\rm{p}}}\right]\frac{\alpha_\Lambda}{2a} \sum_\pm \sum_i (\tilde{\Delta}_i^\pm \tilde{\phi}) \tilde{B}_{i,\pm i}^{(4)} \; ,\label{eq:gaussLATprog}
\end{align}
where $\tilde{\pi}_{\phi}=\dot{\tilde{\phi}}$ and the last equation stands for the lattice version of Gauss's law. 

Similarly, the lattice version of Eqs.~(\ref{eq:friedman1CONT})-(\ref{eq:friedman2CONT}) for the expansion of the universe are simply obtained by considering the volume averages of the different energy contributions,
\begin{align}
    \ddot{\tilde{a}} &= -\left[\frac{f_{*}}{m_{\rm{p}}}\right]^2\frac{a}{3}\left(2\tilde{\rho}^{\rm L}_{\rm K}-\tilde{\rho}^{\rm L}_{\rm V}+\tilde{\rho}^{\rm L}_{\rm EM}\right)\, ,\label{eqn:friedman1LATprog}\\[4pt]
  \tilde{H}^2 &= \left[\frac{f_{*}}{m_{\rm{p}}}\right]^2\frac{1}{3}(\tilde{\rho}^{\rm L}_{\rm K}+\tilde{\rho}^{\rm L}_{\rm G}+\tilde{\rho}^{\rm L}_{\rm V}+\tilde{\rho}^{\rm L}_{\rm EM})\; . \label{eq:friedman2LATprog}
\end{align}
with
\begin{equation}
\begin{gathered}
\tilde{\rho}^{\rm L}_{\rm K} \equiv \frac{1}{2}\left\langle\tilde{\pi}^2_{\phi}\right\rangle_{\rm V}\; , \quad \tilde{\rho}^{\rm L}_{\rm G} \equiv \frac{1}{2a^2}\Big\langle(\tilde{\Delta}^{\hspace{-0.5mm}+}\hspace{-0.5mm}\tilde{\phi})^2\Big\rangle_{\rm V}\; ,
\\[4pt]
\tilde{\rho}^{\rm L}_{\rm V} \equiv \left\langle \tilde{V}(\tilde{\phi})\right\rangle_{\rm V}\;, \quad\tilde{\rho}^{\rm L}_{\rm EM} \equiv \left[\frac{\omega_{*}}{f_{*}}\right]^2\frac{1}{2a^4} \Big\langle a^2  \tilde{E}^2_i + \tilde{B}^2_i\Big\rangle_{\rm V} \;,
\end{gathered}
\label{eqn:energyDensityTermsLatticeProg}
\end{equation}
where $\langle ... \rangle_{\rm V} = \frac{1}{N}\sum_{\textbf{n}}(...)$ represents the volume average over the entire lattice.

This is a hybrid scheme where time is not discretised in the action; instead time discretisation takes place when considering the specific time integrator. This feature makes this scheme naturally adaptable to dynamical equations that include terms that depend on the canonical momenta of the field to be evolved in the kernels. As a consequence, non-symplectic time integrators can be directly applied so as to get an explicit temporal evolution of the system \cite{Figueroa:2024rkr}, which is not the case when fully discretised actions are considered \cite{Cuissa:2018oiw}. In this work, we use a second-order low-storage Runge-Kutta integrator \cite{Carpenter1994Thirdorder2R,Carpenter1994Fourthorder2R,Figueroa:2021iwm} from which we reproduce the continuum limit up to $\mathcal{O}(dx^2, dt^2)$.

Finally, we have used the public package \CLns~\cite{Figueroa:2020rrl,Figueroa:2021yhd}, suitably modified to account for the axial coupling and arbitrary potentials, as the computational framework to implement our lattice formulation.

\subsubsection{Initial conditions}

We follow the initialisation procedure presented in \cite{Figueroa:2024rkr}, where independent initial states can be set for each of the transverse polarisation states of the gauge fields, namely $A^{+}$ and $A^{-}$. In order to accomplish that, one needs to write the cartesian components of the gauge field in the chiral basis:

\begin{equation}\label{eq:gaugeInChiral}
    A_{i} = A^{+}\varepsilon_i^+ + A^{-}\varepsilon_i^- \,, ~~i = 1,2,3\, ,
\end{equation}
where the chiral vectors $\varepsilon_i^\pm(\theta,\varphi)$ are defined as in Eq.~(\ref{eq:chiral_vec}).

We note that even though the initial configuration is completely transversal in the backreaction-less regime, once inhomogeneities start to kick-off, the axial interaction induces longitudinal modes to appear in the gauge sector. Our lattice prescription  naturally captures such longitudinal modes, since all physical degrees of freedom are evolved.

The solution of the gauge field chiral components in the backreaction-less regime and for sufficiently sub-Hubble modes $k\gg(aH)$ is that of the Bunch-Davies solution \cite{Anber:2006xt}. If a simulation were to be initiated under such conditions—where the full set of momenta captured on the lattice is at very sub-Hubble scales—the Bunch-Davies solution must be imposed for every polarisation. Alternatively, one can choose to have a set of modes that are already mildly super-Hubble at the initial stage of the simulation. This implies that, as the peak of the excitation of the tachyonic mode (\ref{eqn:amplifiedAplus}) roughly follows the Hubble scale, part of the initial spectrum is already significantly excited above the vacuum solution, while most UV modes remain in the vacuum solution. 

 \begin{figure}[t!]
	\centering
	\includegraphics[width=0.7\textwidth]{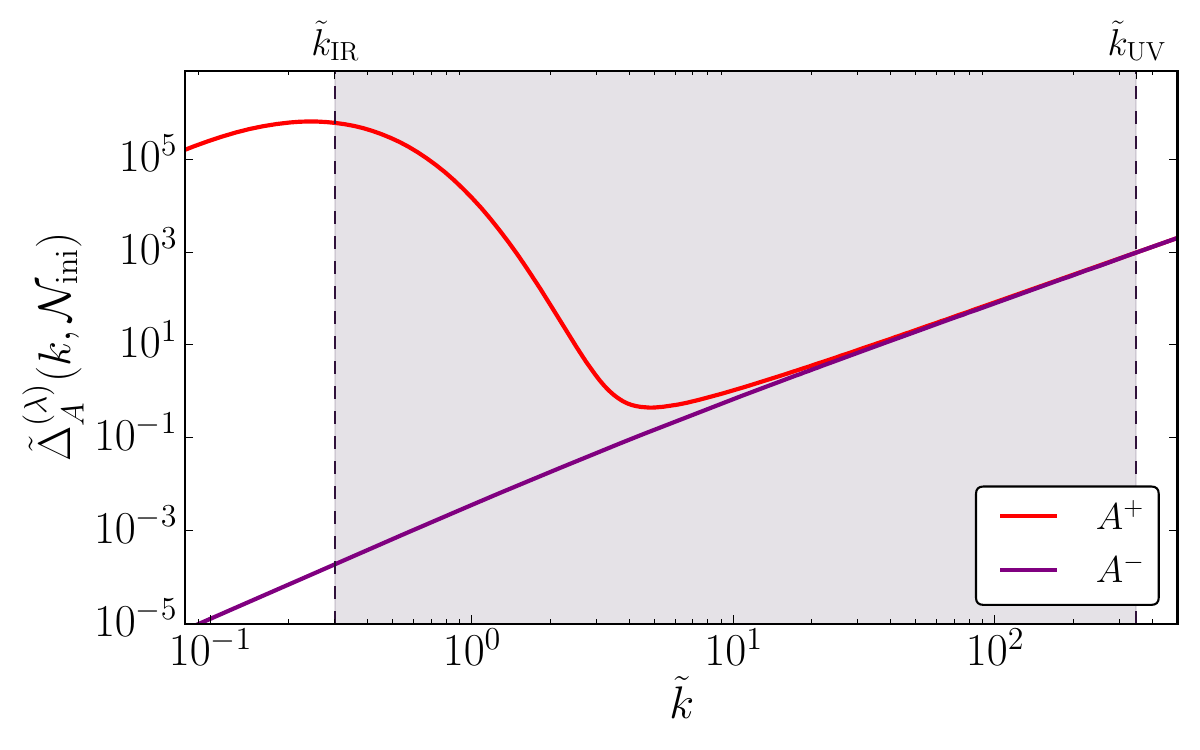}
	\caption{The power spectra of the positive $A^{+}$ (red) and negative $A^{-}$ (purple) helicities at initialisation $\mathcal{N}=\mathcal{N}_{\rm ini}$. The gray shaded band represents the comoving momentum window captured by the lattice, with $k_{\rm IR}\sim (aH)_{\rm ini}$ and $N=1344$ points per dimension. Example obtained using the potential corresponding to natural inflation with $\alpha_{\Lambda}=17.9$.}\label{fig:GaugePS_ini}
\end{figure}

Here we opt to use the second methodology, for which we closely reproduce the steps presented in \cite{Figueroa:2024rkr}. Firstly, the initialisation moment is suitably chosen so that the transition from negligible backreaction to the regime where backreaction is noticeable is performed smoothly and progressively. Similarly, we select lattices with comoving momenta spanning from $k_{\rm IR} \sim (aH)_{\rm ini}$ to $k_{\rm UV} \sim \mathcal{O}(10^2)-\mathcal{O}(10^3) (aH)_{\rm ini}$, depending on the required dynamical range. The exact values are included in Table~\ref{tab:latt_par_1} and vary from case to case. Our previous paper \cite{Figueroa:2024rkr} includes a detailed discussion of this procedure and comparisons with the methodology where the gauge field starts at the vacuum solution. We do not intend to extend that discussion, but here we just note that a special care must be taken in the initialisation methodology we employ in this paper, as there exist a non-trivial momentum-dependent relation between the complex configuration of the Fourier modes of the gauge fields and their canonical momenta---the electric field.

Fig.~\ref{fig:GaugePS_ini} shows an illustrative example of the gauge field configuration we use as the initial condition of the system. We include the initial power spectra of the two helicities of the gauge field $\tilde{\Delta}_A^{(\pm)}(\tilde{k}, \mathcal{N})=\frac{\tilde{k}^3}{2\pi^3}|\tilde{A}^{\pm}(\tilde{k}, \mathcal{N})|^2$. The momentum window typically captured by our lattices is highlighted as a dashed region between $k\in[k_{\rm IR}, k_{\rm UV}]$.

\subsection{Single-field potentials}
\label{subsec:potentials}
The form of the potential plays a crucial role in shaping the inflationary trajectory. Not only that, but the generation of primordial perturbations and gravitational waves also depends strongly on its exact form. Axion inflation has traditionally been studied using the original axion-like natural potentials $V(\phi)\sim 1-\cos(\phi/v)$, and especially the simplest monomial chaotic potentials $V(\phi)\sim\phi^{2}$ (see, for instance, \cite{Barnaby:2011vw,Meerburg:2012id,Cheng:2015oqa,Notari:2016npn,Domcke:2020zez,Gorbar:2021rlt,vonEckardstein:2024tix,Adshead:2015pva,Adshead:2018doq,Adshead:2019igv,Adshead:2019lbr,Cuissa:2018oiw,Domcke:2023tnn,Adshead:2023mvt,Fujita:2025zoa,Sharma:2024nfu,Gorbar:2023zla,Caravano:2022epk,Caravano:2021bfn,Figueroa:2024rkr,Figueroa:2023oxc}).

\begin{table}[h]
	\centering
    {\renewcommand{\arraystretch}{1.25}
\begin{tabular}{|c|c|c||c|c|}
	\hline
Name & $V(\phi)$  & Constants &  $\tilde{V}(\tilde{\phi})$ & Model scales \\
	\hline \hline
\multirow{2}{*}{Monodromy}	& \multirow{2}{*}{$ \mu^3 \left( \sqrt{\phi^2+\phi_c^2} - \phi_c \right)$} & $\phi_c = m_{\rm p}/10$ & \multirow{2}{*}{$\sqrt{\tilde{\phi}^2+1} - 1$} & $f_*=\phi_c$ \\
& & $\mu = 6.0 \cdot 10 ^{-4} m_{\rm p} $ & & $\omega_*=\mu^{3/2}/\sqrt{\phi_c}$  \\
	\hline
\multirow{2}{*}{Starobinsky}	& \multirow{2}{*}{$ V_0 \left( 1 - \exp \left( - \frac{|\phi |}{v} \right) \right)^2 $ }& $v = 10 m_{\rm p}/3$ & \multirow{2}{*}{$ \left( 1 - e^{-|\tilde{\phi} | } \right)^2$ }& $f_*=v$ \\
& &\makecell[c]{ $V_0 = 6.2 \cdot 10^{-10} m_{\rm p}^4$\vspace{0.5mm}} & & $\omega_*=\sqrt{V_0}/v$   \\
	\hline
    \multirow{2}{*}{Hilltop}	& \multirow{2}{*}{$ V_0 \left( 1 - \left( \frac{|\phi|}{v}\right)^4 \right)^2 
$} & $v = 8 m_{\rm p}$ & \multirow{2}{*}{$\left( 1 - \tilde{\phi}^4\right)^2$} &  $f_*=v$  \\
& & \makecell[c]{ $V_0 = 5.7 \cdot 10^{-11} m_{\rm p}^4$ \vspace{0.5mm}} & &$\omega_*=\sqrt{V_0}/v$  \\
	\hline
Chaotic 	& $\frac{1}{2}m^2 \phi^2$ & $m = 6.16\cdot 10^{-6}m_{\rm{p}}$ &$\frac{1}{2}\tilde{\phi}^2$ & $f_*=\omega_*=m$   \\
	\hline
\multirow{2}{*}{Natural}	& \multirow{2}{*}{$V_0 \left( 1 + \cos \frac{\phi}{v} \right)$} & $v = \sqrt{8\pi} m_{\rm p}$ & \multirow{2}{*}{$1 + \cos \tilde{\phi} $} & $f_*=v$ \\
& & \makecell[c]{$V_0 = 5.9\cdot 10^{-10} m_{\rm p}^4$ \vspace{0.5mm}} & & $\omega_*=\sqrt{V_0}/v$ \\
	\hline
\multirow{2}{*}{$\mathbf{\alpha}$-attractor(2)}	& \multirow{2}{*}{$ \frac{\Lambda^4}{2}\tanh^2 \frac{|\phi|}{M}$} & $M=8.79 m_{\rm p}$ & \multirow{2}{*}{$ \frac{1}{2}\tanh^2 |\tilde{\phi}|$} & $f_*=M$ \\
& & $\Lambda = 7.16\cdot 10^{-3} m_{\rm p}  $ & & $\omega_*=\Lambda^2/M$ \\
	\hline
\multirow{2}{*}{$\mathbf{\alpha}$-attractor(4)}	& \multirow{2}{*}{$\frac{\Lambda^4}{4}\tanh^4 \frac{|\phi|}{M}$} & $M=8.22m_{\rm{p}}$ & \multirow{2}{*}{$ \frac{1}{4}\tanh^4 |\tilde{\phi}|$} & $f_*=M$ \\
& & $\Lambda = 8.47\cdot 10^{-3} m_{\rm p}  $ & &  $\omega_*=\Lambda^2/M$\\
	\hline
\end{tabular}
\caption{Definition of the form and scale of the potentials used in the models simulated for this work. We also include $\tilde{V}(\tilde{\phi})$ expressed in program variables and the definition of the typical energy scale $f_{*}$ and fundamental timescale $\omega_{*}$ in each case.}\label{tab:potentials}
}
\end{table}

This work seeks to go beyond those conventional models in axion inflation by exploring a broader class of theoretically well-motivated single-field potentials. This selection has been guided by both theoretical motivations and phenomenological relevance. We aim to encompass a diverse landscape of models, including those rooted in fundamental theories such as supergravity and string theory, as well as phenomenological constructions that capture key features of inflationary dynamics. Some of these potentials, such as the $\alpha$-attractors and the Starobinsky models\footnote{We note that the recent release by the Atacama Cosmology Telescope (ACT) might partially modify the landscape of observationally preferred inflationary potentials \cite{ACT:2025tim,ACT:2025fju}.}, remain among the most favoured by current CMB measurements \cite{Planck:2018jri,BICEP:2021xfz}. Others are less preferred, or even ruled out like the chaotic model. We still include them for comparisons with works in the literature and to allow for a comprehensive investigation into the effects of potential shape and structure on the dynamics of axion inflation during the strong backreaction regime. The Table~\ref{tab:potentials} summarizes the selected models, with the corresponding names and model parameters, specified to match the Planck amplitude at $60$ e-folds before the end of inflation, in the absence of backreaction \cite{Adshead:2019lbr,Figueroa:2024yja}. The table also contains the program version of the potentials, with the corresponding characteristic energy scale ($f_{*}$) and timescale ($\omega_{*}$).

The following list contains a summary of the origin and basic features of our selected models:

\begin{itemize}
   \item \textbf{Chaotic}. Originally developed in \cite{Linde:1983gd}, chaotic inflation represents one of the simplest realizations of inflationary dynamics, characterized by monomial potentials such as $V(\phi) \propto \phi^n$. Although Planck data have disfavoured these models due to their predictions for a large tensor-to-scalar ratio, they have been extensively studied and remain a cornerstone in the historical development of inflationary theory. We include this potential as a benchmark reference, as it was used in our previous works \cite{Figueroa:2023oxc,Figueroa:2024rkr}, to allow direct comparison with the other potentials considered in this paper.
   
    \item \textbf{Natural}. Natural inflation was the first axion-inspired model, proposed in \cite{Freese:1993bc}. The potential has a periodic form and is motivated by spontaneously broken global symmetries. Compatibility with Planck data requires a super-Planckian axion decay constant \cite{Adams:1992bn}, which challenges the model's theoretical consistency. The axial coupling $\phi F\tilde{F}$ was originally proposed in \cite{Anber:2009ua} to reconcile the model with observation, allowing sub-Planckian decay constants.
    
    \item \textbf{Monodromy}. Axion monodromy inflation was proposed in \cite{Silverstein:2008sg} and conforms a set of string-theory motivated models that allow large-field excursions of the inflaton. It leads to potentials that often take forms such as linear, fractional power-law, or modulated shapes.
    
    \item \textbf{Starobinsky}. It is also known as $R^2$ inflation \cite{Starobinsky:1980te}, and arises from the addition of a $R^2$ term to the Einstein-Hilbert action, which dominates during inflation. Through a conformal transformation, this theory can be recast as a standard Einstein theory with a canonical scalar field and an exponentially flat potential \cite{Maeda:1988ab,Whitt:1984pd}.

    \item $\mathbf{\alpha}$\textbf{-attractor(}$\mathbf{2n}$\textbf{)}. $\alpha$-attractors form a class of super-conformal inflationary models \cite{Kallosh:2013hoa,Kallosh:2013yoa}. They are characterized by the parameter $\alpha$, which controls the curvature of the inflaton’s field space and governs the universality of the inflationary predictions. For small $\alpha$, they exhibit a universal behaviour, with predictions for $n_{\rm{s}}$ and $r$ that are compatible with observations. We have considered two particular potentials of this class from a subset called T-models, which take the form of even powers of the hyperbolic tangent. Some other $\alpha$-attractors are classified as E-models \cite{Kallosh:2013hoa,Kallosh:2013yoa,Galante:2014ifa, Kallosh:2015lwa}, their form depends on negative exponentials, and Starobinsky models are a particular case within this group.

    \item \textbf{Hilltop}. This class of inflation, originally proposed in \cite{Boubekeur:2005zm}, refers generically to all models for which inflation occurs near to a local maximum of the potential, typically leading to small-field dynamics. These models are characterized by a potential that is flat near the origin and steepens as the field rolls away. Here we have studied a particular form proposed in \cite{Kallosh:2019jnl}.
\end{itemize}

 \begin{figure}[h]
	\centering
	\includegraphics[width=0.8\textwidth]{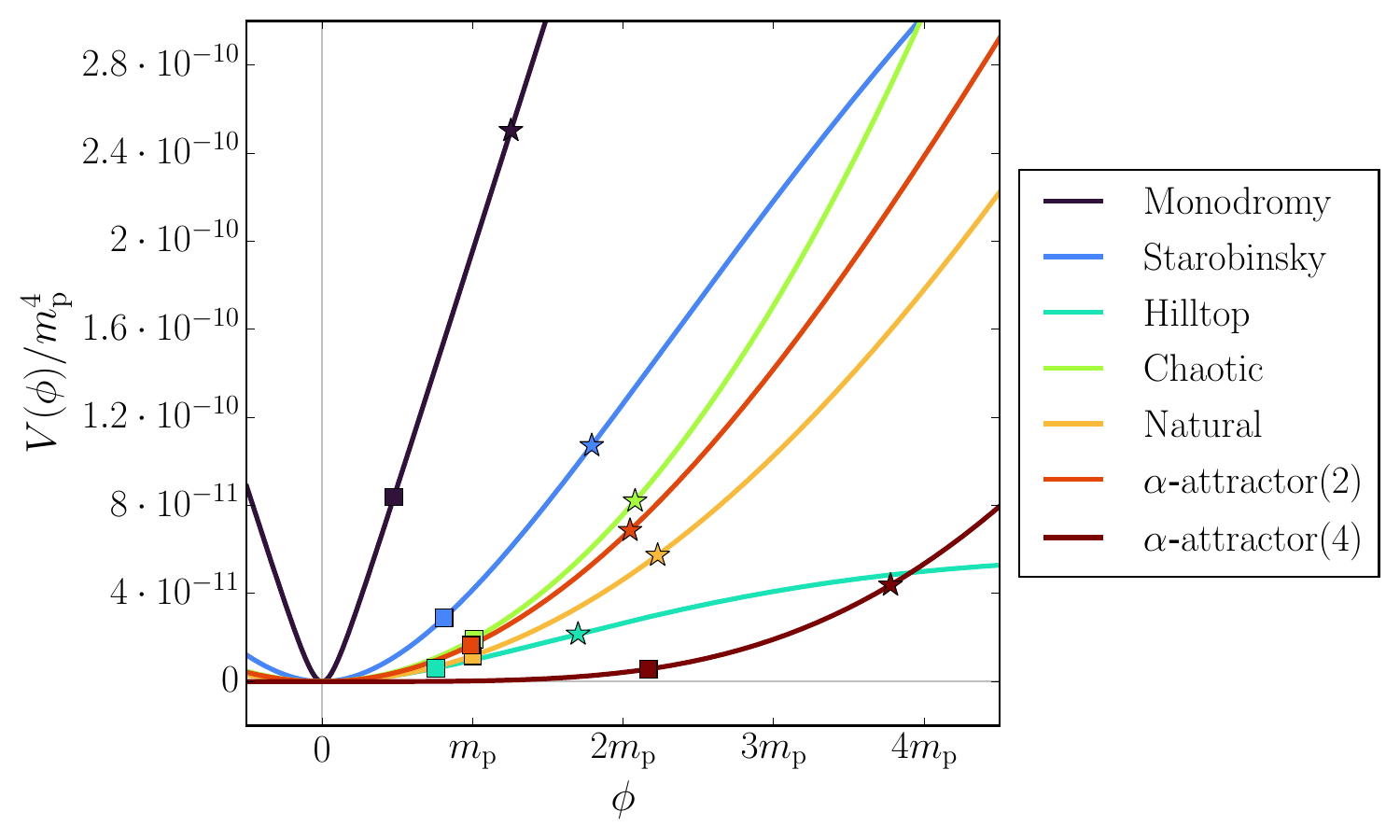}
	\caption{Representation of all the potentials considered around the relevant inflaton field scales on which the gauge field backreaction starts to be considerable. The stars correspond to $\mathcal{N}_{\rm{ini}}$ for each potential for the strongest $\alpha_{\Lambda}$ simulated in each case, compiled in Tab.~\ref{tab:latt_par_1}. The squares are the inflaton field values at $\mathcal{N}=0$ related to the end of inflation in the backreaction-less approach.}
    \label{fig:Im_pot}
\end{figure}

Figure~\ref{fig:Im_pot} shows the comparison of the shape of all potentials considered around the scales probed by our simulations. Note that those with the minimum away from $\phi=0$ have been shifted for comparative purposes.  We indicate the value of the scalar field at initialisation using stars, for the strongest coupling considered, or, equivalently, with the earliest initialisation time (see the next section for specific values of the strongest couplings and corresponding initialisation moments). Similarly, we use squares to indicate the end of backreaction-less inflation, \textit{i.e.} $\mathcal{N}=0$. 

\begin{figure}[h]
	\centering
	\includegraphics[width=\textwidth]{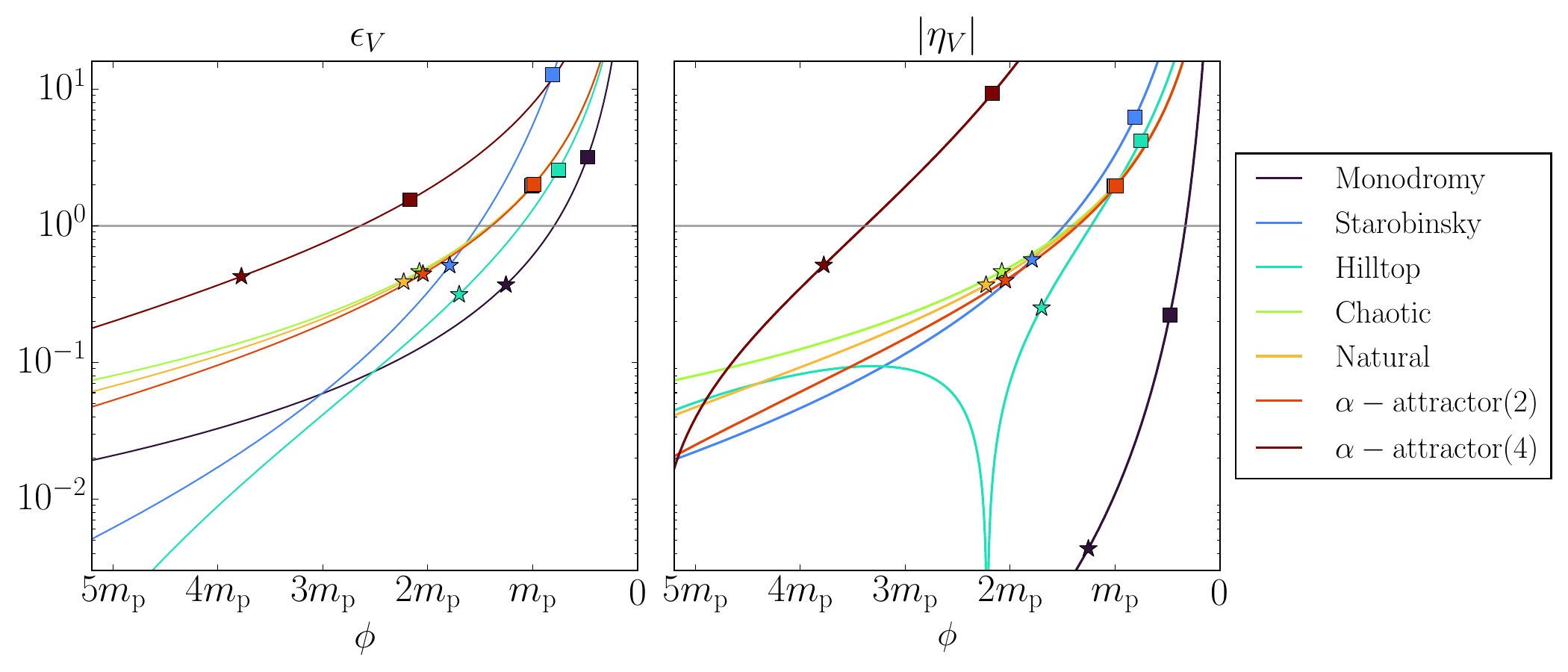}
	\caption{Slow-roll parameters $\epsilon_V$ (left) and absolute value of $\eta_V$  (right) represented for each potential studied. The square and starts in each potential represent the same e-folds as in Fig.~\ref{fig:Im_pot}.}\label{fig:Im_SR}
\end{figure}

In order to gain more insight in the comparison of the potentials, we include additionally the evolution of the slow-roll parameters $\epsilon_V$ and $\eta_V$ in Fig.~\ref{fig:Im_SR}. These are defined as
\begin{equation}
\epsilon_V = \frac{m_{\rm p}^2}{2}\left( \frac{V'}{V} \right)^2, \qquad \eta_V = m_{\rm p}^2\frac{V''}{V} ,
\end{equation}
where prime denotes differentiation with respect $\phi$. They provide information about the scale-independent steepness and curvature of the potential.

The figure already reveals notable similarities among certain models, particularly those that exhibit a quadratic-like profile, namely, Natural, Chaotic, and $\alpha$-attractor(2), as they align remarkably well in both relevant parameters within the range probed by our simulations, \textit{i.e.} between the markers indicated by the star and square in Fig.~\ref{fig:Im_pot}. Thus, we expect them to produce comparable effects. Nevertheless, this preliminary analysis does not provide sufficient grounds for anticipating analogous behaviours from the remaining potentials, whose characteristics diverge more significantly within the same parameter space.

\subsection{Simulations and parameters}
\label{subsec:ICandRunParams}
\begin{table}[htbp]
{\renewcommand{\arraystretch}{1.0}
\newcolumntype{M}[1]{>{\centering\arraybackslash}m{#1}}
	\centering
	$$\begin{array}{|M{2.8cm}|c||M{1.8cm}|M{1.8cm}|M{1.8cm}|M{1.8cm}|}
		\hline  
		\multirow{2}{*}{$ \text{Potential} $} & \multirow{2}{*}{$ \alpha_\Lambda$}& \multirow{2}{*}{$ \tilde{k}_{\rm IR}$} &\multirow{2}{*}{$  \tilde{k}_{\rm UV} $}& \multirow{2}{*}{$ \mathcal{N}_{\rm ini} $}&\multirow{2}{*}{$  N$}\\ & & & & & \\
        \hline
		\rm\multirow{2}{*}{ Monodromy } & \multirow{4}{*}{$\begin{array}{l c}
        \text{SBR-lim.}: &  18.9 \\
        \text{+5\%}: &  19.8 \\
        \text{+10\%}: &  20.8 \\
        \text{+15\%}: &  21.7 \\
        \text{+20\%}: &  22.7 \\
        \end{array}$}
          & 0.1 & 41.569 & -0.6 & 480 \\
		& & 0.1 & 41.569 & -0.7 & 480\\
		& & 0.1 & 41.569 & -0.8 & 480\\
		& & 0.1 & 62.354 & -0.9 & 720\\
        & & 0.1 & 96.995 & -0.9 & 1120\\
        \hline
		\rm\multirow{2}{*}{Starobinsky} & \multirow{4}{*}{$\begin{array}{l c}
        \text{SBR-lim.}: &  17.8 \\
        \text{+5\%}: &  18.7 \\
        \text{+10\%}: &  19.6 \\
        \text{+15\%}: &  20.5 \\
        \text{+20\%}: &  21.4 \\
        \end{array}$}
          & 0.2 & 83.138 & -0.9 & 480 \\
		& & 0.2 & 83.138 & -1.0 & 480\\
		& & 0.3 & 187.061 & -0.9 & 720\\
		& & 0.3 & 299.298 & -1.0 & 1152\\
        & & 0.3 & 299.298 & -1.0 & 1152\\
        \hline
        \rm\multirow{2}{*}{Hilltop} & \multirow{4}{*}{$\begin{array}{l c}
        \text{SBR-lim.}: &  20.8 \\
        \text{+5\%}: &  21.8 \\
        \text{+10\%}: &  22.9 \\
        \text{+15\%}: &  23.9 \\
        \text{+20\%}: &  25.0 \\
        \end{array}$}
          & 0.9 & 374.123 & -0.7 & 480 \\
		& & 0.9 & 374.123 & -0.8 & 480\\
		& & 0.9 & 561.184 & -0.9 & 720\\
		& & 0.9 & 897.895 & -1.0 & 1152\\
        & & 0.9 & 872.954 & -1.0 & 1120\\
		\hline	
        \rm\multirow{2}{*}{ Chaotic} & \multirow{4}{*}{$\begin{array}{l c}
        \text{SBR-lim.}: &  14.25 \\
        \text{+5\%}: &  15.0 \\
        \text{+10\%}: &  15.7 \\
        \text{+15\%}: &  16.4 \\
        \text{+20\%}: &  17.1 \\
        \end{array}$}
          & 0.3 & 124.708 & -0.9 & 480 \\
		& & 0.3 & 124.708 & -1.0 & 480\\
		& & 0.4 & 249.415 & -1.0 & 720\\
		& & 0.4 & 399.065 & -1.0 & 1152\\
        & & 0.4 & 387.979 & -1.0 & 1120\\
		\hline
		\rm\multirow{2}{*}{  Natural} & \multirow{4}{*}{$\begin{array}{l c}
        \text{SBR-lim.}: &  14.9 \\
        \text{+5\%}: &  15.6 \\
        \text{+10\%}: &  16.4 \\
        \text{+15\%}: &  17.1 \\
        \text{+20\%}: &  17.9 \\
        \end{array}$}
          & 0.2 & 83.138 & -1.0 & 480 \\
		& & 0.2 & 83.138 & -1.2 & 480\\
		& & 0.3 & 187.061 & -1.2 & 720\\
		& & 0.3 & 299.298 & -1.2 & 1152\\
        & & 0.3 & 349.181 & -1.2 & 1344\\
        \hline
		\rm\multirow{2}{*}{$\alpha $ -attractor(2)} & \multirow{4}{*}{$\begin{array}{l c}
        \text{SBR-lim.}: &  15.2 \\
        \text{+5\%}: &  16.0 \\
        \text{+10\%}: &  16.7 \\
        \text{+15\%}: &  17.5 \\
        \text{+20\%}: &  18.2 \\
        \end{array}$}
          & 0.3 & 124.798 & -0.9 & 480 \\
		& & 0.3 & 187.061 & -1.0 & 480\\
		& & 0.4 & 249.415 & -1.0 & 720\\
		& & 0.4 & 399.065 & -1.0 & 1152\\
        & & 0.4 & 387.979 & -1.0 & 1120\\
        \hline
		\rm\multirow{2}{*}{$\alpha $ -attractor(4)} & \multirow{4}{*}{$\begin{array}{l c}
        \text{SBR-lim.}: &  12.6 \\
        \text{+5\%}: &  13.2 \\
        \text{+10\%}: &  13.9 \\
        \text{+15\%}: &  14.5 \\
        \text{+20\%}: &  15.1 \\
        \end{array}$}
          & 0.1 & 41.569 & -1.0 & 480 \\
		& & 0.1 & 99.766 & -1.1 & 1152\\
		& & 0.1 & 133.022 & -1.3 & 1536\\
		& & 0.1 & 133.022 & -1.5 & 1536\\
        & & 0.1 & 155.192 & -1.5 & 1792\\
        \hline
	\end{array}$$
	\caption{Compilation of the lattice parameters used for each potential. The simulated $\alpha_{\Lambda}$ couplings correspond to the SBR-limit and increments of $+5\%$, $+10\%$, $+15\%$, and $+20\%$.}	
	\label{tab:latt_par_1}	}
\end{table}	

This section includes the set of parameters, with their corresponding numerical values, used in the simulations presented in this work. We aimed to explore the dynamics of the system for couplings that belong to the strong backreation regime (SBR). In order to do so, first we identify the minimum value of $\alpha_\Lambda$ that yields, without excursions out of inflation, non-zero extra e-folds as compared with the backreaction-less case. We follow the same strategy as in the right panel of Fig.~7 of \cite{Figueroa:2024rkr} to find this specific value, which we call the SBR-limit. We then extend the range of coupling values by adding fixed increments of $+5\%$ with respect to that minimum value up to $+20\%$.

Table~\ref{tab:latt_par_1} includes the values of the parameters needed to set the lattice and initial conditions for each case. All of them have been determined to fulfil the features of the initialisation described in the previous section (see Fig.~\ref{fig:GaugePS_ini}) \cite{Figueroa:2024rkr}. $N$ represents the number of sites per dimension of the lattices, $\tilde{k}_{\rm IR}$ is the dimensionless wavenumber associated to the longest wavelength captured by the lattice, $\tilde{k}_{\rm UV}$ that of the smallest wavelength and $\mathcal{N}_{\rm ini}$ is the initialisation moment that corresponds to number of e-folds before the end of inflation in the backreaction-less limit which we set at $\mathcal{N}=0$.


\section{Results}
\label{sec:results}
In this section we report on the results of the analysis of the strong backreaction regime for our set of potentials and coupling constants. We divide this section into two subsections. In Sec.~\ref{subsec:results_dynamics}, we first present the outcomes from simulations with different potentials and perform a systematic comparison between them. Then, in Sec.~\ref{subsec:results_homo}, we extend this comparison to include the results using the homogeneous backreaction approach to assess the range of validity of such technique. 

We mainly use three fiducial potentials to present the results: Natural, Monodromy, and the $\alpha$-attractor$(4)$. As we will see along the section, the remaining potentials mostly resemble one of these three, making them suitable for illustrating the observed trends.

\subsection{Dynamics in the strong backreaction regime}
\label{subsec:results_dynamics}

The dynamics in the strong backreaction regime is characterised by a defining feature: a considerable lengthening of the inflationary period. This additional inflationary stage is sustained by a continuous exponential energy transference from the inflaton potential to the gauge sector. As a consequence, the electromagnetic contribution to the energy density dominates more and more over the kinetic part of the inflaton, which also shows a non-negligible inhomogeneity generation in the form of gradient energy density. This new energy hierarchy is the responsible for the emergence of extra e-folds in inflation. Inflation proceeds until the electromagnetic contribution becomes comparable to that of the potential, and as a result, inflation ends with a Universe completely reheated. We denominated this new regime as \textit{electromagnetic slow-roll} of axion inflation in our previous works \cite{Figueroa:2023oxc, Figueroa:2024rkr} (see also \cite{Sharma:2024nfu} for similar analyses), where it was extensively studied for the case of the chaotic potential. In this work we extend that analysis to the set of potentials presented in the previous section. 

\begin{figure}[t!]
	\centering
    \begin{subfigure}{\textwidth}     
    \includegraphics[width=0.96\textwidth]{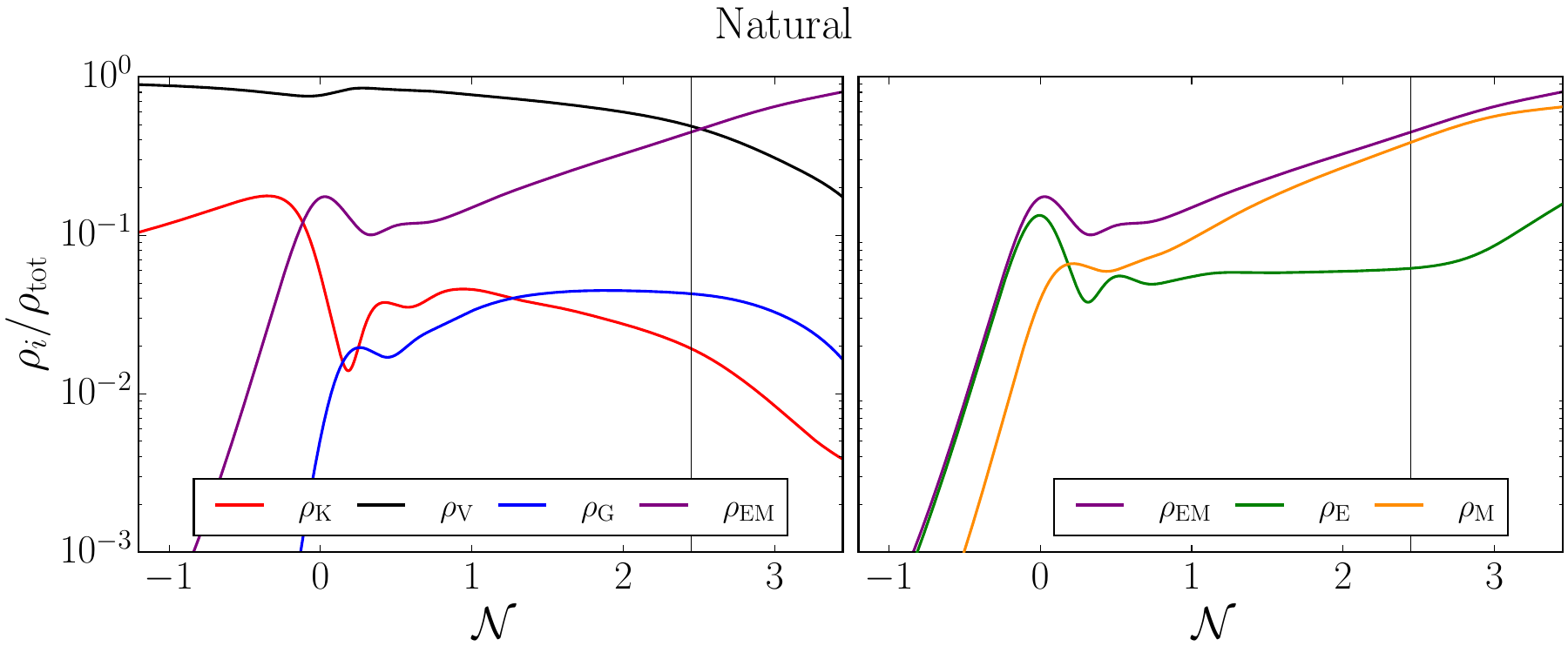} 
    \end{subfigure}
    \begin{subfigure}{\textwidth}
      \includegraphics[width=0.96\textwidth]{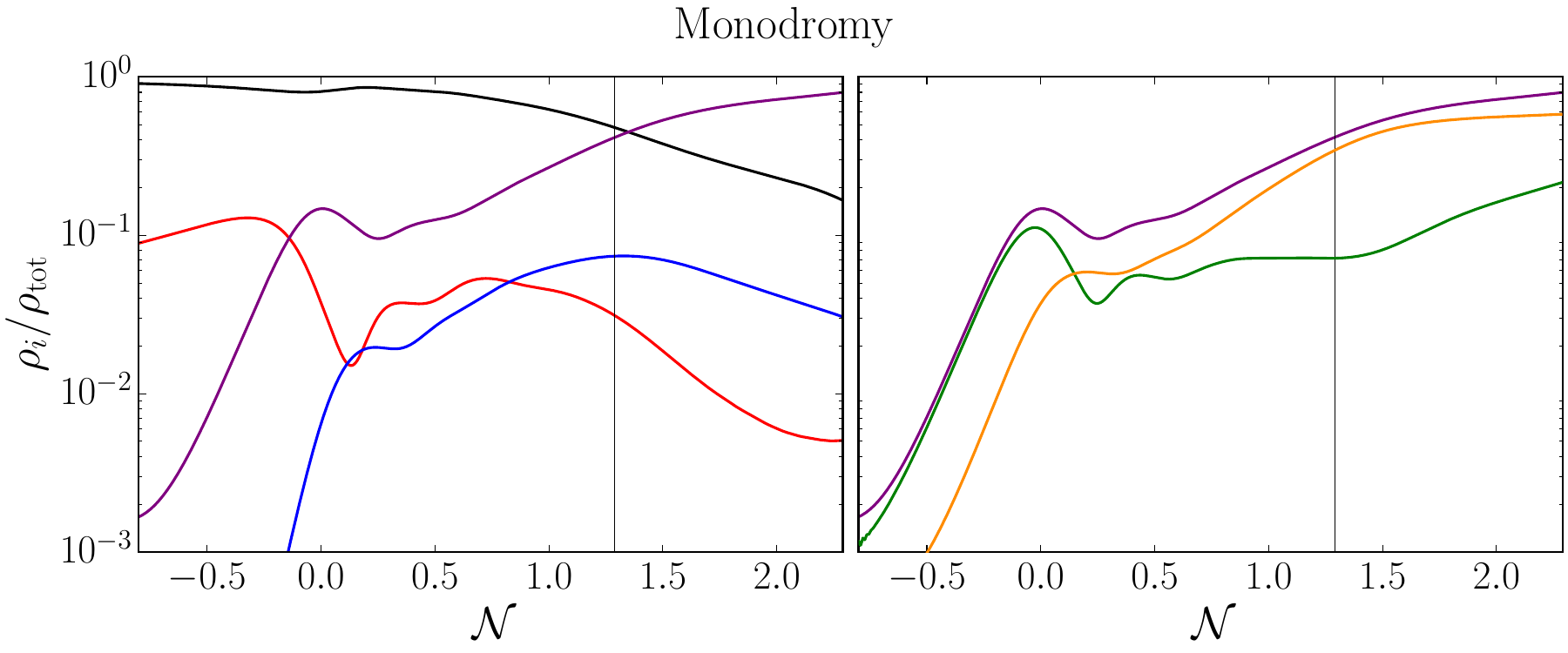} 
    \end{subfigure}
\begin{subfigure}{\textwidth}
      \includegraphics[width=0.96\textwidth]{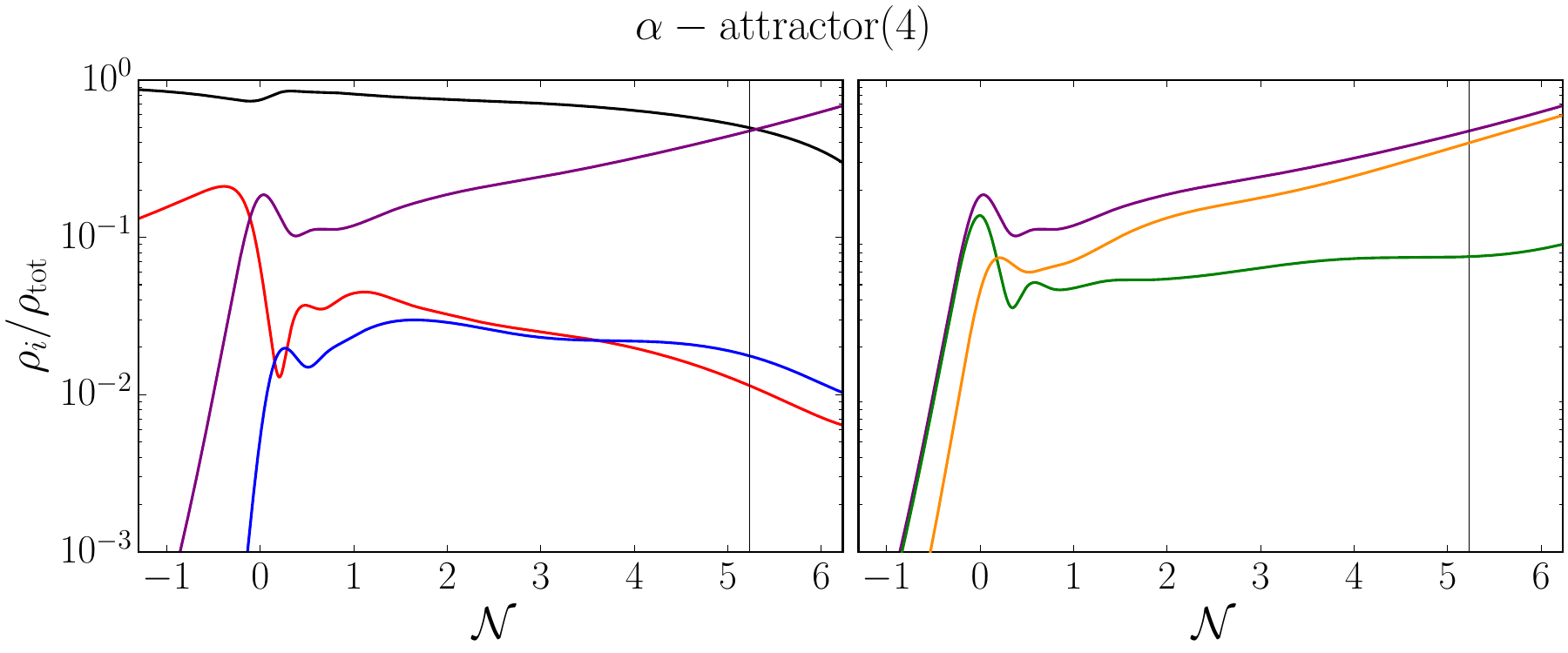} 
    \end{subfigure}
  
    \caption{\textit{Left}: Evolution of the different energy densities normalised to the total for the fiducial potentials: Natural, Monodromy and $\alpha$-attractor(4). We show the potential (black), kinetic (red), gradient (blue) and electromagnetic (purple) energy densities. \textit{Right}: Evolution of the normalised electromagnetic energy density (purple) and its magnetic (orange) and electric (green) contributions. In both panels the end of inflation is indicated by a vertical line. The coupling used corresponds to SBR$+10\%$.}
    \label{fig:energy_densities}
\end{figure}

In the left panels of Fig.~\ref{fig:energy_densities}, we show for the three fiducial potentials with couplings corresponding to SBR+$10\%$ the evolution of the different contributions to the energy density of the system normalised to the total contribution: potential (black), kinetic (red), gradient (blue) and electromagnetic (purple). Aside from specific quantitative values, we observe that there is a good qualitative agreement between different models, and also with the results presented in \cite{Figueroa:2023oxc, Figueroa:2024rkr} and summarized in the previous paragraph. Specifically, after the onset of backreaction, there is a longer period of inflation (with the end indicated using a vertical line) due to the electromagnetic slow-roll. All cases also show that during that period the gradient energy becomes comparable to the kinetic one, eventually surpasses it and that it can acquire approximately $10\%$ of the total energy budget depending on the model.

In the right panels of Fig.~\ref{fig:energy_densities} we show the evolution of the total electromagnetic energy density (purple lines) and its decomposition into magnetic (orange lines) and electric (green lines) components for each fiducial potential. Another important feature we observe in the figure --- common across all types of inflationary potentials --- is that in the electromagnetic slow-roll regime, it is the magnetic energy density contribution that dominates over the electric one. In all three cases, the system starts with a dominant $\rho_E$, which decreases once the backreaction becomes significant and is eventually overtaken by the magnetic contribution. It is this latter component that governs the dynamics during the extra e-folds of inflation. This behaviour is fully analogous to what was found for the quadratic potential in our previous works \cite{Figueroa:2023oxc,Figueroa:2024rkr}, indicating that it is intrinsic to the model in the strong backreaction regime, regardless of the inflationary potential.

The UV sensitivity of the axion inflation model also reported in \cite{Figueroa:2023oxc,Figueroa:2024rkr}, and therefore, the need to cover a sufficiently wide dynamical range so as not to fake the physical results are features also present in the full set of potentials. For this reason, in all cases, we ensure that the inflation parameter $\epsilon_H$—which we define in the next paragraph—has converged at the end of inflation within $1\%$ when varying the dynamical range by increasing the resolution, except for the $\alpha$-attractor(4) where we reach $3-4\%$ for the largest couplings. As it will be detailed below, this last potential exhibits the largest increase in extra inflationary e-folds with respect to the coupling, and therefore displays the most extreme UV sensitivity, which hits our computational limits. In order to reach convergence, larger lattices were required as the value of $\alpha_{\Lambda}$ was increased, as shown in Tab.~\ref{tab:latt_par_1}.

There are also evident quantitative differences, mainly in the duration of the electromagnetic slow-roll period, and hence in the lengthening of inflation. Table~\ref{tab:endInflation} includes this information, where the number of extra e-folds in inflation, with respect to the backreaction-less case, are represented as $\mathcal{N}_{\rm end}$.\footnote{We note that $\mathcal{N}_{\rm end}$ is equivalent to ``$\Delta\mathcal{N}_{\rm br}$'' used in \cite{Figueroa:2023oxc,Figueroa:2024rkr}} We compute this values by following 

\begin{equation}\label{eq:epsilonH}
    \epsilon_{H}=-\frac{\dot{H}}{H^2}=1+\frac{2\rho_{\rm{K}}-\rho_{\rm{V}}+2\rho_{\rm{EM}}}{\rho_{\rm{tot}}}\;,
\end{equation}
which allows to identify the end of inflation $\mathcal{N}=\mathcal{N}_{\rm{end}}$ at $\epsilon_H = 1$. We show the evolution of this global parameter in the next section.

Using Tab.~\ref{tab:endInflation}, we identify $\alpha$-attractor(4) and Monodromy inflation as opposite extreme examples. The former is clearly the model with most extra e-folds, ranging from $\mathcal{N}_{\rm end}\sim 3.5$ for the SBR-limit coupling to $\mathcal{N}_{\rm end}\sim9.2$ for the largest $\alpha_{\Lambda}$. This means that it requires considerably more e-folds to finish inflation than our baseline chaotic model of \cite{Figueroa:2023oxc, Figueroa:2024rkr}. Monodromy, on the contrary, is in the opposite side, predicting fairly low number of extra e-folds for the same relative values of $\alpha_\Lambda$, with $\mathcal{N}_{\rm{end}}\sim2.2$ for $+20\%$. The rest of the cases predict intermediate and similar amounts of extra inflation. In particular, as we have already envisaged in the previous section in the analysis of the slow-roll parameters, Natural, Chaotic and $\alpha$-attractor(2) lead to very similar behaviour, with $1.4\lesssim\mathcal{N}_{\rm end}\lesssim3.6$ for the coupling range we consider.  

\begin{table}[t!]
{\renewcommand{\arraystretch}{1.0}
\begin{tabular}{|c|c|c|}
		\hline  
		Potential & $ \alpha_\Lambda$ & $ \mathcal{N}_{\rm end}$ \\
        \hline
		\rm\multirow{5}{*}{ Monodromy } & \multirow{5}{*}{$\begin{array}{l c}
        \text{SBR-lim.}: &  18.9 \\
        \text{+5\%}: &  19.8 \\
        \text{+10\%}: &  20.8 \\
        \text{+15\%}: &  21.7 \\
        \text{+20\%}: &  22.7 \\
        \end{array}$} & 0.614 \\
        & & 0.924 \\
        & & 1.290 \\
        & & 1.698 \\
        & & 2.184 \\
        \hline
		\rm\multirow{5}{*}{Starobinsky} & \multirow{5}{*}{$\begin{array}{l c}
        \text{SBR-lim.}: &  17.8 \\
        \text{+5\%}: &  18.7 \\
        \text{+10\%}: &  19.6 \\
        \text{+15\%}: &  20.5 \\
        \text{+20\%}: &  21.4 \\
        \end{array}$} & 1.209 \\
        & & 1.515 \\
        & & 1.896 \\
        & & 2.437 \\
        & & 2.908 \\
        \hline
        \rm\multirow{5}{*}{Hilltop} & \multirow{5}{*}{$\begin{array}{l c}
        \text{SBR-lim.}: &  20.8 \\
        \text{+5\%}: &  21.8 \\
        \text{+10\%}: &  22.9 \\
        \text{+15\%}: &  23.9 \\
        \text{+20\%}: &  25.0 \\
        \end{array}$} & 1.213 \\
        & & 1.484 \\
        & & 1.823 \\
        & & 2.180 \\
        & & 2.678 \\
		\hline	
        \rm\multirow{5}{*}{ Chaotic} & \multirow{5}{*}{$\begin{array}{l c}
        \text{SBR-lim.}: &  14.25 \\
        \text{+5\%}: &  15.0 \\
        \text{+10\%}: &  15.7 \\
        \text{+15\%}: &  16.4 \\
        \text{+20\%}: &  17.1 \\
        \end{array}$} & 1.406 \\
        & & 1.824 \\
        & & 2.368 \\
        & & 2.935 \\
        & & 3.535 \\
		\hline
\end{tabular}
\hfill
\begin{tabular}{|c|c|c|}
        \hline  
		Potential & $ \alpha_\Lambda$ & $ \mathcal{N}_{\rm end}$ \\
        \hline
\rm\multirow{5}{*}{  Natural} & \multirow{5}{*}{$\begin{array}{l c}
        \text{SBR-lim.}: &  14.9 \\
        \text{+5\%}: &  15.6 \\
        \text{+10\%}: &  16.4 \\
        \text{+15\%}: &  17.1 \\
        \text{+20\%}: &  17.9 \\
        \end{array}$} & 1.468 \\
        & & 1.871 \\
        & & 2.449 \\
        & & 3.031 \\
        & & 3.726 \\
        \hline
		\rm\multirow{5}{*}{$\alpha $ -attractor(2)} & \multirow{5}{*}{$\begin{array}{l c}
        \text{SBR-lim.}: &  15.2 \\
        \text{+5\%}: &  16.0 \\
        \text{+10\%}: &  16.7 \\
        \text{+15\%}: &  17.5 \\
        \text{+20\%}: &  18.2 \\
        \end{array}$} & 1.481 \\
        & & 1.927 \\
        & & 2.403 \\
        & & 3.002 \\
        & & 3.554 \\
        \hline
		\rm\multirow{5}{*}{$\alpha $ -attractor(4)} & \multirow{5}{*}{$\begin{array}{l c}
        \text{SBR-lim.}: &  12.6 \\
        \text{+5\%}: &  13.2 \\
        \text{+10\%}: &  13.9 \\
        \text{+15\%}: &  14.5 \\
        \text{+20\%}: &  15.1 \\
        \end{array}$} & 3.536 \\
        & & 4.288 \\
        & & 5.293 \\
        & & 6.410 \\
        & & 8.198 \\
        \hline
\end{tabular}
}
	\caption{Numerical values of the coupling constant and the delay of the end of inflation $\mathcal{N}_{\rm end}$.}\label{tab:endInflation}
\end{table}

\begin{figure}[t!]
	\centering
	\includegraphics[width=\textwidth]{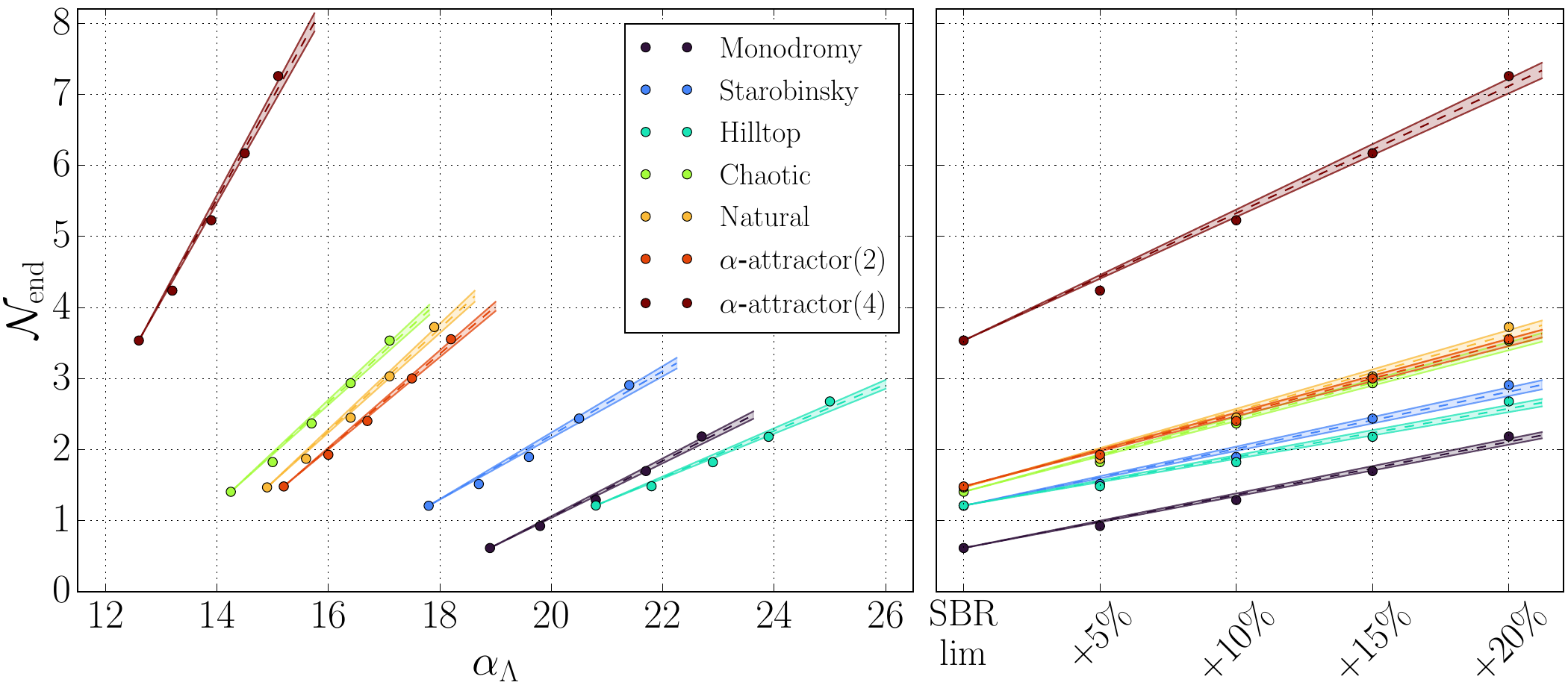}
	\caption{Delay of the end of inflation $\mathcal{N}_{\rm end}$ for each simulated model. The left panel shows the explicit dependence with the coupling constant, while in the right panel it is represented versus the coupling constants normalised with respect to the value of $\alpha_\Lambda$ at the SBR-limit. The dashed coloured lines on the left represent the linear fit (\ref{eq:linearFit}) with the shaded region corresponding to the $1\sigma$ errors. The parameters of the linear fits are given in Tab.~\ref{tab:latt_par_1}.}\label{fig:Im_compare}
\end{figure}

Fig.~\ref{fig:Im_compare} includes the graphic representation of the results. The left panel shows $\mathcal{N}_{\rm end}$ against $\alpha_\Lambda$. A clear monotonic, seemingly linear, relation between the delay of the end of inflation and the coupling value can be observed. This agrees with the findings for the chaotic model in \cite{Figueroa:2024rkr}, and we find to be a common feature of all the potentials considered. The dashed line with the shaded band represents a linear fit to the points (see below for more details). In addition, the comparison for different potentials shows that there is an apparent anti-correlation between $\mathcal{N}_{\rm end}$ and the absolute values of $\alpha_\Lambda$ used in SBR for each case, \textit{i.e.} more (less) extra e-folds are predicted for the potentials that require smaller (larger) values of the coupling. The right panel shows an alternative comparison where on the horizontal axis we use relative couplings computed by normalising with respect to the value of $\alpha_\Lambda$ at the SBR-limit. This figure shows that the relative increases for different potentials share a similar pattern, growing roughly in parallel trajectories. Furthermore, it evidences more clearly the clustering of different models, specially for Natural, Chaotic and $\alpha$-attractor(2) which lie on top of each other.

As we did in \cite{Figueroa:2024rkr}, we propose two possible fits to interpolate the results:
\begin{align}
    \text{Linear: } \quad & \mathcal{N}_{\rm end} = m(\alpha_\Lambda - \alpha_{\Lambda, 0}) + \mathcal{N}_{{\rm end},0}\;,\label{eq:linearFit}\\
    \text{Power - law: } \quad & \mathcal{N}_{\rm end} = b(\alpha_\Lambda - \alpha_{\Lambda, 0})^a + \mathcal{N}_{{\rm end},0} \;,\label{eq:powerlawFit}
\end{align}
where $ \alpha_{\Lambda, 0} $ and $ \mathcal{N}_{{\rm end},0} $ correspond to the coupling constant and extra e-folds of inflation of the SBR-limit. The first fit corresponds to a linear function of one parameter, while the second fit depends on two parameters and accounts for deviations from the linear behaviour. The results for the fit parameters and their errors are shown in Tab. \ref{tab:fitttingParams}. 

\begin{table}[t!]
\begin{center}
{\renewcommand{\arraystretch}{1.25}
\begin{tabular}{|c||c c||c c|c c|}
\hline
 Potential & $\hat{m}$ & $\sigma_m$ & $\hat{a}$ & $\sigma_a$ & $\hat{b}$ & $\sigma_b$ \\
\hline
Monodromy & 0.396 & 0.011 & 1.177 & 0.029 & 0.325 & 0.01 \\
Starobinsky & 0.451 & 0.017 & 1.243 & 0.047 & 0.348 & 0.019 \\
Hilltop & 0.328 & 0.012 & 1.242 & 0.045 & 0.244 & 0.014 \\
Chaotic & 0.719 & 0.020 & 1.1825 & 0.0051 & 0.6176 & 0.0029 \\
Natural & 0.722 & 0.021 & 1.1931 & 0.0061 & 0.6089 & 0.0036 \\
$\alpha$-attractor(2) & 0.667 & 0.017 & 1.1656 & 0.0020 & 0.5759 & 0.0011 \\
$\alpha$-attractor(4) & 1.422 & 0.041 & 1.194 & 0.020 & 1.240 & 0.019 \\
\hline
\end{tabular}
}
\caption{Mean values and errors of the parameters of the fit functions of Eqs.~(\ref{eq:linearFit})-(\ref{eq:powerlawFit}).} \label{tab:fitttingParams}
\end{center}
\end{table}

As anticipated previously, we include a graphic representation of the linear fit~\eqref{eq:linearFit} in Fig.~\ref{fig:Im_compare}, with dashed lines for the mean and shaded bands representing the error of the fitting. The figure shows that the linear fit gives a fairly good fit to the data. The grouping of different models is also evident in the slopes for the linear fit. The potential in the lowest range of $\alpha_\Lambda$, $\alpha$-attractor(4),  has the steepest slope. For the potentials in the intermediate range, the slope gets progressively milder for Chaotic,\footnote{The value of $m$—not to be confused with mass of the axion—obtained in this work for the chaotic potential is close to, yet not compatible to that reported in \cite{Figueroa:2024rkr}. We note that the higher couplings considered for the fit in that work yield a slightly larger value.} Natural and $\alpha$-attractor(2), respectively. Meanwhile, in the highest range of the couplings, conformed by Monodromy, Starobinsky and Hilltop, the slope is flatter than in the previous potentials. This would suggest an approximate clustering of the potentials around the fiducial ones, defining three groups. On another note, the power law fits indicate that the data seem to prefer a slightly concave form rather than a perfectly linear growth, with a common power index of $a\sim1.2$ for all potentials. This shows that, although the slopes differ within the selected set, the growth pattern appears to be universal to the choice of the potential. 

As a final remark, we note that the similarity groups are qualitatively reflected in the slow-roll parameters $\epsilon_{\rm{V}}$ and $\eta_{\rm{V}}$ of Fig.~\ref{fig:Im_SR}. However, we argue that these parameters can only provide rough estimates, since the observed differences in the system's dynamics are comparatively more significant.

\subsection{Comparison with homogeneous methods}
\label{subsec:results_homo}

Our previous works \cite{Figueroa:2023oxc,Figueroa:2024rkr} demonstrated the necessity to respect the truly local nature of the backreaction in order to correctly capture the physics of the model. The homogeneous backreaction scheme, on the contrary, albeit computationally more efficient, failed to provide an accurate description, specially when inhomogeneities (gradients) become sufficiently relevant. In order to assess whether this situation is transversal and potential-independent, or whether there could be a better agreement under certain circumstances, we perform a systematic comparison of the dynamics depicted by the two aforementioned methods for the different potentials considered in Tab.~\ref{tab:potentials}. We use an adapted version of our code to account for the homogeneous backreaction, as shown and validated in \cite{Figueroa:2024rkr}.

\begin{figure}[t!]
	\centering
    \begin{subfigure}{\textwidth}
        \includegraphics[width=1\textwidth]{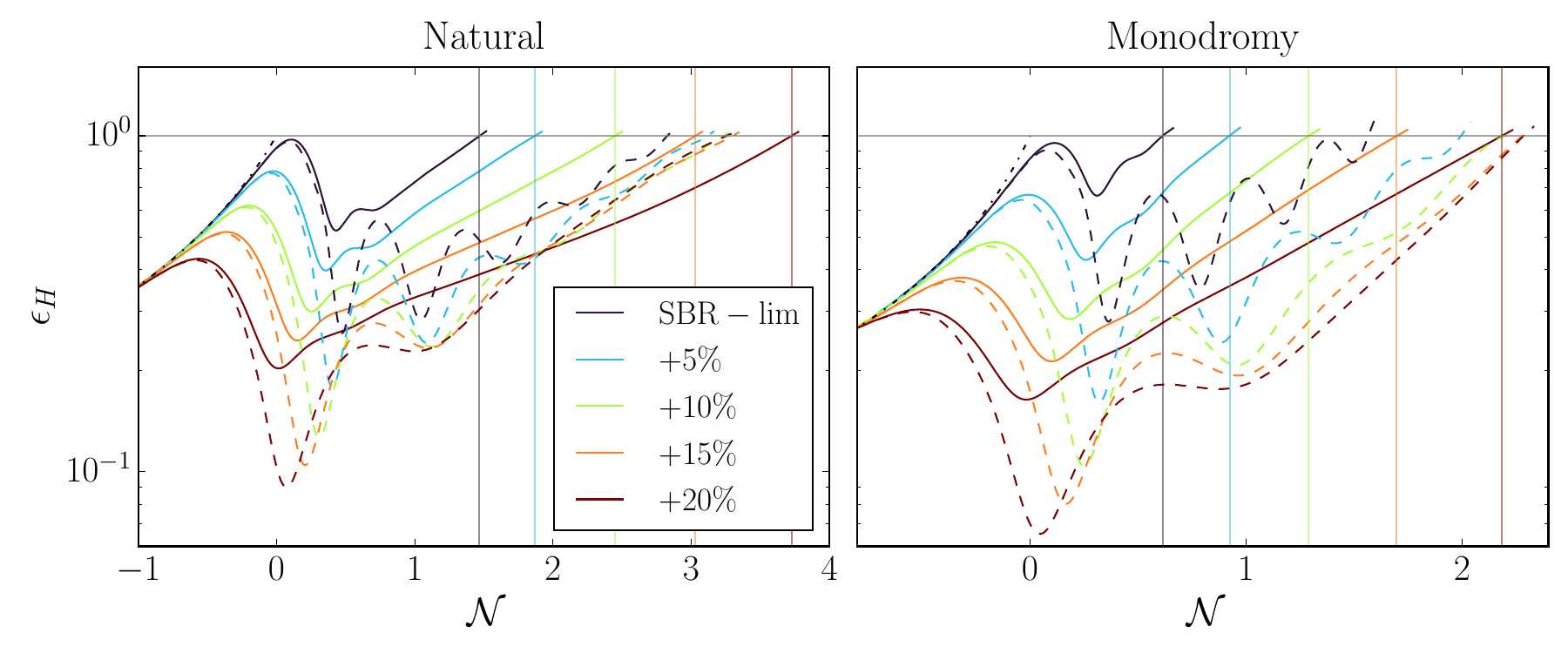} 
    \end{subfigure}
    \begin{subfigure}{0.572\textwidth}
        \includegraphics[width=1\textwidth]{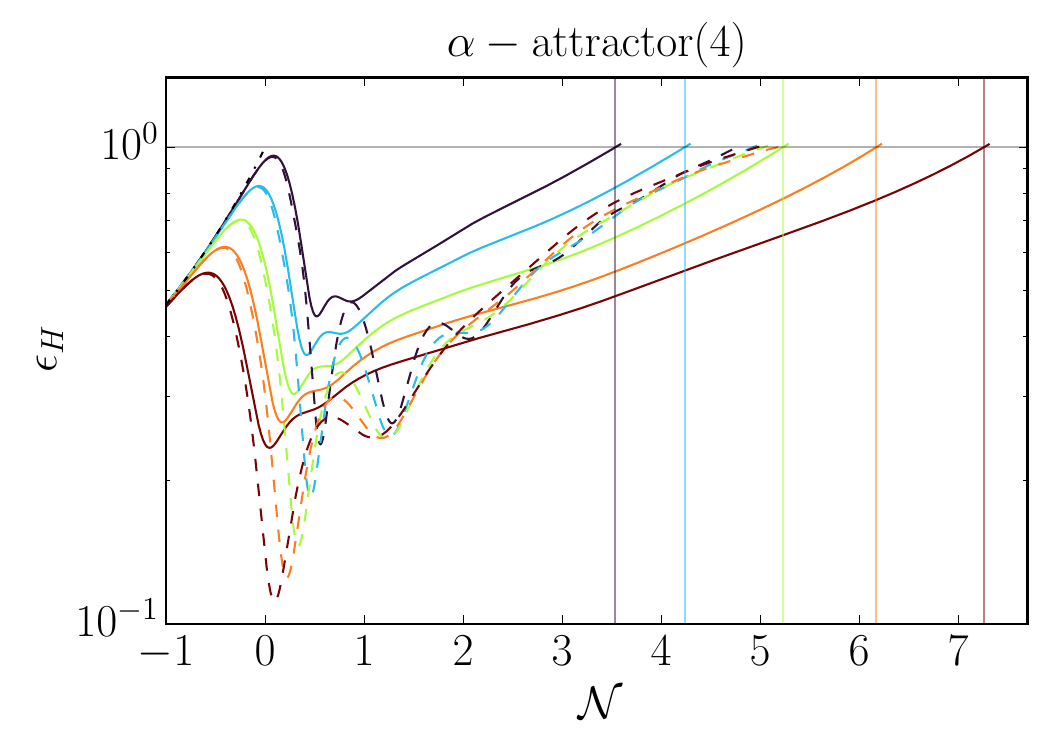} 
    \end{subfigure}
    \caption{Evolution of $\epsilon_H$, for the three different fiducial potentials and for all couplings, corresponding to lattice simulations of the local backreaction (solid lines) and homogeneous backreaction methods (dashed lines), as well as the backreaction-less case (dash-dotted black line). The value of $\alpha_\Lambda$ that corresponds to the SBR-limit is represented in black, whereas the values increased by $+5\%$ are shown in blue,  $+10\%$ in green, $+15\%$ in orange and $+20\%$ in brown. The vertical lines mark the condition $\epsilon_H=1$ for each model, according to the local backreaction results.}
    \label{fig:epsilonH_fiducialPots}
\end{figure}

In Fig.~\ref{fig:epsilonH_fiducialPots}, we compare the evolution of $\epsilon_H$ for the local backreaction (solid lines) and homogeneous backreaction (dashed lines). We use the same three representative potentials: Natural in the top left panel, Monodromy in the top right panel, and $\alpha$-attractor$(4)$ in the bottom panel. We include all couplings that have been simulated: SBR-limit (black) and increases of $+5\%$ (blue), $+10\%$ (green), $+15\%$ (orange) and $+20\%$ (brown). As a reference, we also include the backreaction-less evolution of $\epsilon_H$ (dash-dotted black line) and indicate with coloured vertical lines the end of inflation for each case in the local backreaction simulations.

\begin{figure}[t!]
	\centering
    \begin{subfigure}{0.49\textwidth}
        \includegraphics[width=1\textwidth]{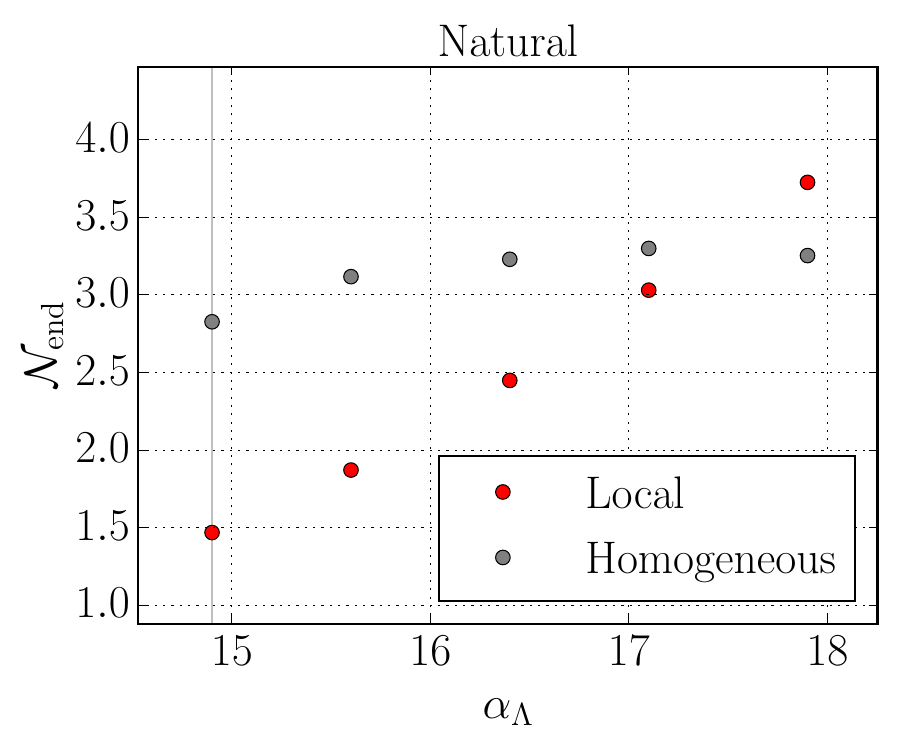} 
    \end{subfigure}
    \begin{subfigure}{0.49\textwidth}
        \includegraphics[width=1\textwidth]{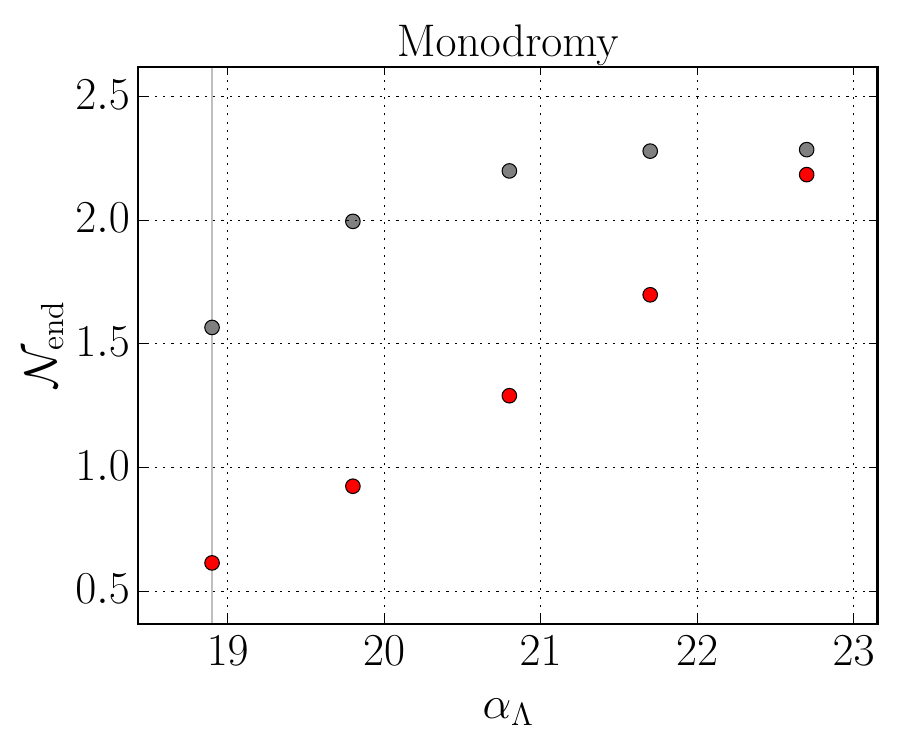} 
    \end{subfigure}
    \begin{subfigure}{0.49\textwidth}
        \includegraphics[width=1\textwidth]{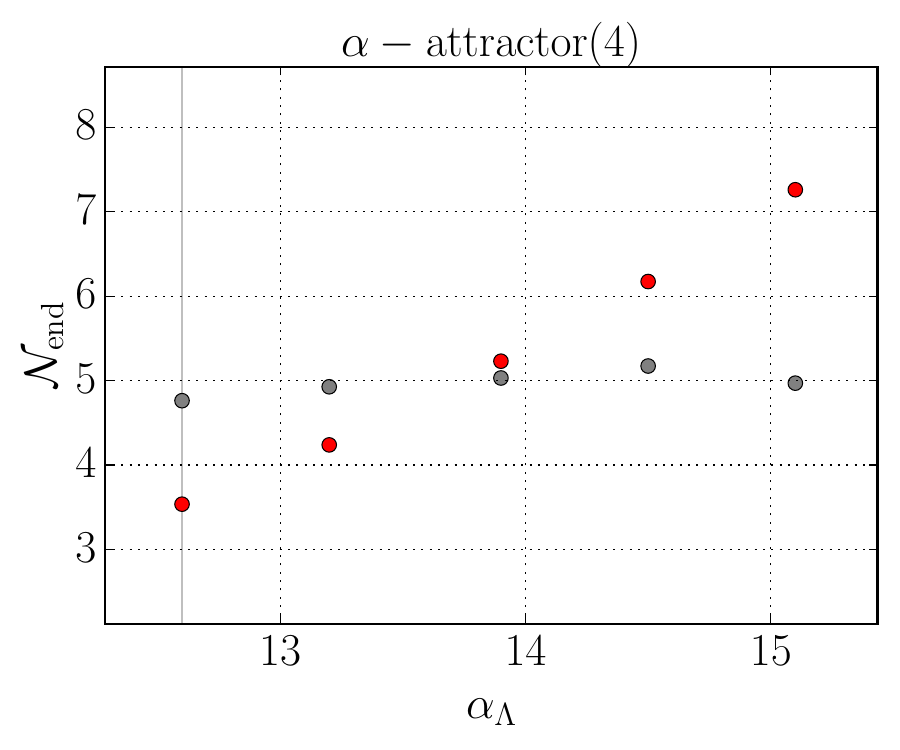} 
    \end{subfigure}
    \caption{Delay of the end of inflation $\mathcal{N}_{\rm end}$ for three fiducial potentials with respect to the value of the coupling constant $\alpha_\Lambda$, corresponding to the local backreaction approach (red dots) and the homogeneous backreaction approach (grey dots). The grey vertical solid lines account for the SBR-limit coupling for each potential.}
    \label{fig:fittingPlotsFiducialPots}
\end{figure}

Both methods share the same trajectories until the backreaction becomes noticeable, generating a localized bump in $\epsilon_H$. In the subsequent evolution, the trajectories go up until $\epsilon_H=1$ is reached and inflation finishes. However, the detailed outcomes of both methods differ significantly. Approximately half an e-fold after the bump, though a bit earlier for Monodromy, both methods start to diverge. The homogeneous approach shows its typical oscillations from this point on, which generates more bumps than the primordial one. We observe that this trend is generic to all potentials considered (also to those four not shown in the figure), although they seem to be slightly attenuated for the biggest couplings. What the local backreaction method shows, in turn, is that there is only one bump in the evolution, the primordial one. After this moment, the electromagnetic slow-roll regime starts, where $\epsilon_H$ grows smoothly. Interestingly, the minimum of $\epsilon_H$ after the first bump is always more pronounced in the homogeneous case, which indicates that the friction that the backreaction of the gauge fields exerts on the inflaton has a greater effect in the expansion than in the local backreaction case. This just reflects the presence of a new energy balance due to the generation of inhomogeneities in the inflaton (see Fig.~\ref{fig:energy_densities}).

\begin{figure}[t!]
	\centering
	\begin{subfigure}{0.49\textwidth}
        \includegraphics[width=1\textwidth]{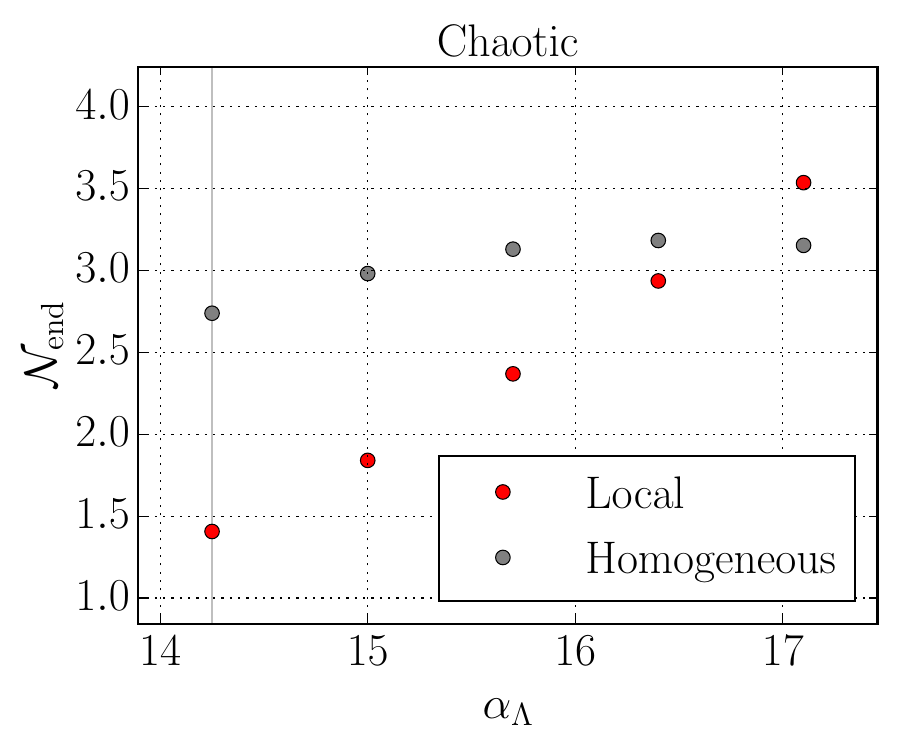} 
    \end{subfigure}
    \begin{subfigure}{0.49\textwidth}
        \includegraphics[width=1\textwidth]{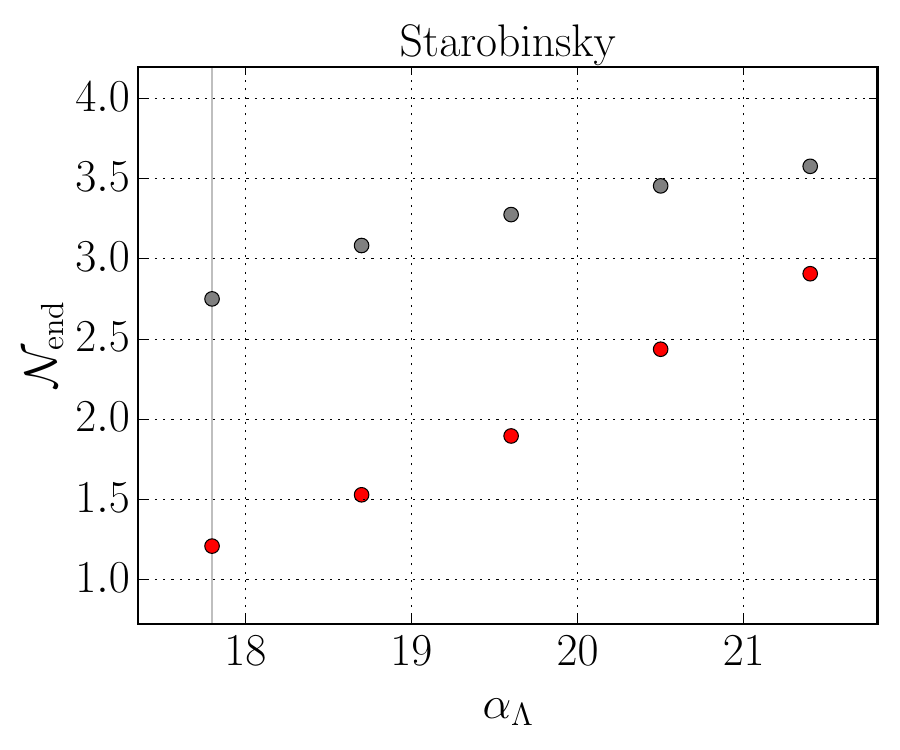} 
    \end{subfigure}
    \begin{subfigure}{0.49\textwidth}
        \includegraphics[width=1\textwidth]{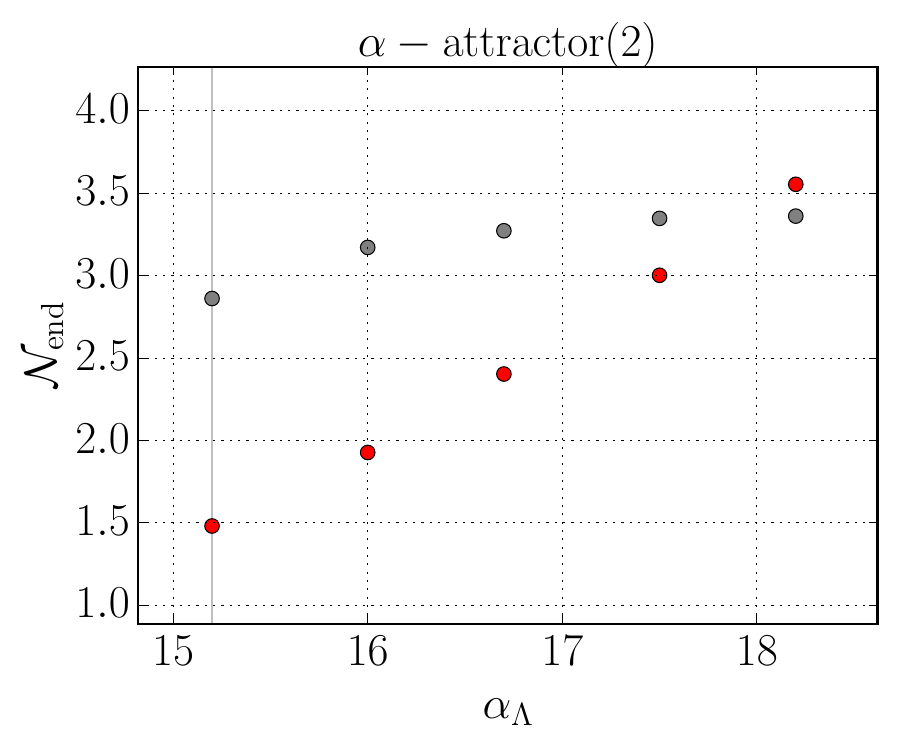} 
    \end{subfigure}
    \begin{subfigure}{0.49\textwidth}
        \includegraphics[width=1\textwidth]{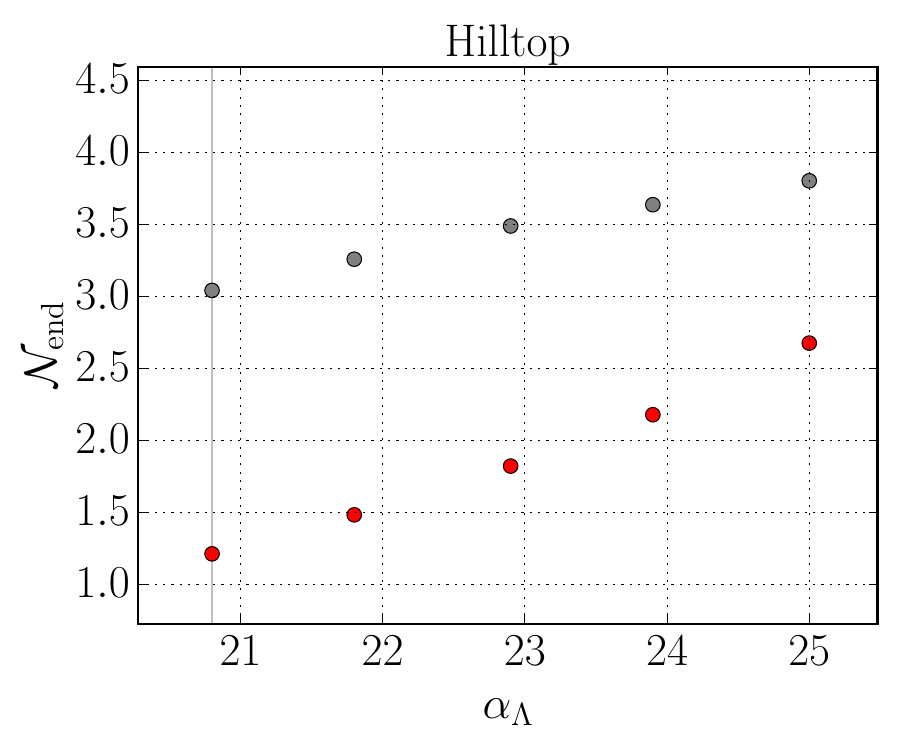} 
    \end{subfigure}
	\caption{Equivalent figure to Fig.~\ref{fig:fittingPlotsFiducialPots} for the rest of the potentials: Chaotic, Starobinsky, $\alpha$-attractor(2) and Hilltop.}
    \label{fig:fittingPlotsRestPots}
\end{figure}

Another considerable difference is the spread of the extra numbers of e-folds required to reach the end of inflation. In the local backreaction case, as indicated in the previous section and for the range of couplings considered, there is a monotonic, almost linear, growth of the duration of inflation with respect to $\alpha_\Lambda$. However, the growth of the homogeneous approach is not that obvious. In fact, even though the onset of backreaction is reached at different moments for each coupling, the dynamics seems to cluster the trajectories, and the number of e-folds required to reach the end of inflation converges to a window of $\sim 0.5-1$ e-folds. As a general rule, the homogeneous scheme predicts more e-folds in inflation for the smallest couplings and fewer e-folds for the largest ones. This is not the case for Monodromy, Starobinsky and Hilltop inflation (see Fig.~\ref{fig:fittingPlotsRestPots}) where the homogeneous scheme yields to more e-folds for all considered couplings. However, given the trend in the local backreaction, we believe that for larger couplings, this could be reverted. We base our estimate on the behaviour we have seen in \cite{Figueroa:2023oxc,Figueroa:2024rkr}  for the chaotic potential, where the linear trend still holds for larger couplings.

An additional way to visualize this difference is to compare $\mathcal{N}_{\rm{end}}$ respect to $\alpha_{\Lambda}$ for both methods. We show this comparison in Figs.~\ref{fig:fittingPlotsFiducialPots} and \ref{fig:fittingPlotsRestPots}, where we include our fiducial potentials, and the rest, respectively. We use red dots for local backreaction and grey dots for homogeneous backreaction, and the vertical grey solid line marks the SBR-limit.

In contrast to the approximately linear growth in the local backreaction scheme, no such behaviour is observed for the homogeneous approach. For all three potentials in Fig.~\ref{fig:fittingPlotsFiducialPots}, the value of $\mathcal{N}_{\rm{end}}$ varies by less than 1 e-fold within the range, with the most noticeable relative change occurring in the Monodromy potential. In all cases, the dependence resembles a concave parabola. In fact, for $\alpha$-attractor(4) at $+20\%$, the value of $\mathcal{N}_{\rm{end}}$ is comparable to that obtained for the coupling at the SBR-limit, and it seems that further increases in coupling will lead to a decrease in $\mathcal{N}_{\rm{end}}$. This behaviour might suggest that beyond a certain potential dependent value, $\mathcal{N}_{\rm{end}}$ might take values equal to or even smaller than those measured at the SBR-limit in the homogeneous backreaction method.

Finally, for completeness, we include the comparison for the rest of the potentials, Chaotic, Starobinsky, $\alpha$-attractor$(2)$ and Hilltop, in Fig.~\ref{fig:fittingPlotsRestPots}; we use the same convention as in Fig.~\ref{fig:fittingPlotsFiducialPots} for the markers and colours. In all four cases, the trend is similar to one of the three fiducial potentials in Fig.~\ref{fig:fittingPlotsFiducialPots}. In fact, we find that the results for the Chaotic and $\alpha$-attractor$(2)$ potentials resemble those of the Natural inflation potential, while the Starobinsky and Hilltop potentials show a similar behaviour to that observed for the Monodromy potential. This is not a new observation, as in the previous section we already defined the same three groups based on the slope of the linear fit.

All in all, we find no indication of convergence between the local and homogeneous methods.


\section{Discussion}

The Chern-Simons coupling between an axion-like inflaton and an Abelian U(1) gauge sector gives rise to a variety of intriguing phenomena and has drawn considerable attention, not only from the theoretical early-universe cosmology community, but also from current and forthcoming cutting-edge observatories. This interaction, however, is in principle independent of the potential responsible for driving inflation, which should be externally specified. The true form of the real inflationary potential remains unknown, and this has motivated this work. Here, we extend our previous studies of the non-linear regime in axion inflation beyond the case of a simple quadratic potential \cite{Figueroa:2023oxc,Figueroa:2024rkr}, and perform a systematic analysis across a diverse set of models to better characterize the dynamics of axion inflation and asses the role of the potential on it.

On a large-scale view of the results, we find that the novel features described in \cite{Figueroa:2023oxc,Figueroa:2024rkr} are qualitatively universal across all potentials considered. In particular, the functional form of the evolution of $\epsilon_H$ is essentially the same for all cases, showing no oscillatory behaviour. We observe that the inflationary period increases with the coupling in a quasi-linear manner, and this quasi-linearity is remarkably similar across different potentials. The dynamics of the different energy contributions also follow an analogous pattern in all cases, with none deviating from the expected electromagnetic slow-roll regime, where the magnetic component dominates over the electric one. Moreover, the system displays UV sensitivity, which requires a large dynamical range to correctly capture the physics of the strongest couplings. Based on these observations, we conclude that these key features are universal in $\phi F\tilde{F}$ Abelian axion inflation and are independent of the specific form of the inflation potential.

Additionally, we have tested the range of validity of the homogeneous backreaction approach. Across a wide range of scenarios—including variations in the steepness of the potential, the duration of inflation, and the magnitude of gradient energy contributions—it consistently diverges from the predictions of the local backreaction framework. These persistent discrepancies underscore the limitations of the homogeneous treatment and the critical role of local inhomogeneities in the backreaction dynamics. This demonstrates that in the region of the SBR probed by our simulations, the use of the local backreaction is imperative. One may wonder whether for larger couplings than those considered here, and hence deeper inside inflation, it would be possible to reach a point where the homogeneous backreaction becomes reconcilable with the local framework. We note, however, that larger values of the couplings were explored in \cite{Figueroa:2023oxc,Figueroa:2024rkr} for the chaotic case, reaching up to $+30\%$ above the coupling at the SBR-limit, and no trend towards convergence was observed. While we cannot rule out the possibility of such reconciliation for significantly higher couplings, our findings for the variety of potentials examined here indicate that it appears unlikely.

On the observational side, the results of this work, together with previous studies \cite{Figueroa:2023oxc,Figueroa:2024rkr}, show that reliable and precise phenomenological predictions must be accompanied by a case-by-case study of the effect of the fully inhomogeneous backreaction on the system's dynamics. Of special relevance is the amount of extra duration of inflation, as some of the analysed potentials, and possibly others not considered in this work, lead to a considerable lengthening of inflation. This might affect potentially observable scales and CMB-relevant scales for instance.

We conclude with some reflective remarks. Even though our works have unveiled the true dynamics of the model for a specially relevant parameter space, it might appear that have also diffused the general picture; the level of uncertainty seems higher than what had been suggested from the previous existing literature. We believe, however, that they have raised the key relevant questions and outlined the new challenges to be pursued from now on. In particular, determining $\mathcal{N}_{\rm{end}}$ in detail for larger couplings to assess whether the linear growth still holds, and if not, understanding what kind of dynamics takes over beyond the electromagnetic slow-roll regime. The role of metric perturbations and their impact on the matter sector during the strong backreaction regime is another issue to be considered. Lattice techniques provide a robust and reliable numerical framework for this purpose, and we plan to address these questions in the near future.


\acknowledgments

We thank Daniel G. Figueroa and Jon Urrestilla for their collaboration during the initial stages of the project and for their valuable comments on the manuscript. A special thanks to Alexandros Papageorgiou for carefully reading the manuscript and spotting a couple of typos in Table~\ref{tab:potentials} and Fig.~\ref{fig:Im_pot}. This work has been done under the support from Eusko Jaurlaritza (IT1628-22) and by the PID2021-123703NB-C21 grant funded by MCIN/AEI/10.13039/501100011033/ and by ERDF; ``A way of making Europe''. In particular, CLM and AU gratefully acknowledge the support from the University of the Basque Country grants PIF24/194 and PIF20/151, respectively. This work has been possible thanks to the computing infrastructure of the Solaris cluster at the University of the Basque Country, UPV/EHU and the Hyperion cluster from the DIPC Supercomputing Center.

\bibliographystyle{JHEP}
\bibliography{AxInfl_Potentials}

\providecommand{\href}[2]{#2}\begingroup\raggedright\begin{thebibliography}{10}

\bibitem{Martin:2013tda}
J.~Martin, C.~Ringeval and V.~Vennin, \emph{{Encyclop\ae{}dia Inflationaris}:
  {Opiparous Edition}},
  \href{https://doi.org/10.1016/j.dark.2024.101653}{\emph{Phys. Dark Univ.}
  {\bfseries 5-6} (2014) 75} [\href{https://arxiv.org/abs/1303.3787}{{\ttfamily
  1303.3787}}].

\bibitem{Planck:2018jri}
{\scshape Planck} collaboration, \emph{{Planck 2018 results. X. Constraints on
  inflation}}, \href{https://doi.org/10.1051/0004-6361/201833887}{\emph{Astron.
  Astrophys.} {\bfseries 641} (2020) A10}
  [\href{https://arxiv.org/abs/1807.06211}{{\ttfamily 1807.06211}}].

\bibitem{Lyth:1998xn}
D.H.~Lyth and A.~Riotto, \emph{{Particle physics models of inflation and the
  cosmological density perturbation}},
  \href{https://doi.org/10.1016/S0370-1573(98)00128-8}{\emph{Phys. Rept.}
  {\bfseries 314} (1999) 1}
  [\href{https://arxiv.org/abs/hep-ph/9807278}{{\ttfamily hep-ph/9807278}}].

\bibitem{Baumann:2009ds}
D.~Baumann, \emph{{Inflation}},  in \emph{{Theoretical Advanced Study Institute
  in Elementary Particle Physics}: {Physics of the Large and the Small}},
  pp.~523--686, 2011, \href{https://doi.org/10.1142/9789814327183_0010}{DOI}
  [\href{https://arxiv.org/abs/0907.5424}{{\ttfamily 0907.5424}}].

\bibitem{Pajer:2013fsa}
E.~Pajer and M.~Peloso, \emph{{A review of Axion Inflation in the era of
  Planck}}, \href{https://doi.org/10.1088/0264-9381/30/21/214002}{\emph{Class.
  Quant. Grav.} {\bfseries 30} (2013) 214002}
  [\href{https://arxiv.org/abs/1305.3557}{{\ttfamily 1305.3557}}].

\bibitem{Baumann:2014nda}
D.~Baumann and L.~McAllister, \emph{{Inflation and String Theory}}, Cambridge
  Monographs on Mathematical Physics, Cambridge University Press (5, 2015),
  \href{https://doi.org/10.1017/CBO9781316105733}{10.1017/CBO9781316105733},
  [\href{https://arxiv.org/abs/1404.2601}{{\ttfamily 1404.2601}}].

\bibitem{Freese:1990rb}
K.~Freese, J.A.~Frieman and A.V.~Olinto, \emph{{Natural inflation with pseudo -
  Nambu-Goldstone bosons}},
  \href{https://doi.org/10.1103/PhysRevLett.65.3233}{\emph{Phys. Rev. Lett.}
  {\bfseries 65} (1990) 3233}.

\bibitem{Adams:1992bn}
F.C.~Adams, J.R.~Bond, K.~Freese, J.A.~Frieman and A.V.~Olinto, \emph{{Natural
  inflation: Particle physics models, power law spectra for large scale
  structure, and constraints from COBE}},
  \href{https://doi.org/10.1103/PhysRevD.47.426}{\emph{Phys. Rev. D} {\bfseries
  47} (1993) 426} [\href{https://arxiv.org/abs/hep-ph/9207245}{{\ttfamily
  hep-ph/9207245}}].

\bibitem{Anber:2006xt}
M.M.~Anber and L.~Sorbo, \emph{{N-flationary magnetic fields}},
  \href{https://doi.org/10.1088/1475-7516/2006/10/018}{\emph{JCAP} {\bfseries
  10} (2006) 018} [\href{https://arxiv.org/abs/astro-ph/0606534}{{\ttfamily
  astro-ph/0606534}}].

\bibitem{Anber:2009ua}
M.M.~Anber and L.~Sorbo, \emph{{Naturally inflating on steep potentials through
  electromagnetic dissipation}},
  \href{https://doi.org/10.1103/PhysRevD.81.043534}{\emph{Phys. Rev. D}
  {\bfseries 81} (2010) 043534}
  [\href{https://arxiv.org/abs/0908.4089}{{\ttfamily 0908.4089}}].

\bibitem{Turner:1987vd}
M.S.~Turner and L.M.~Widrow, \emph{{Gravitational Production of Scalar
  Particles in Inflationary Universe Models}},
  \href{https://doi.org/10.1103/PhysRevD.37.3428}{\emph{Phys. Rev. D}
  {\bfseries 37} (1988) 3428}.

\bibitem{Garretson:1992vt}
W.D.~Garretson, G.B.~Field and S.M.~Carroll, \emph{{Primordial magnetic fields
  from pseudoGoldstone bosons}},
  \href{https://doi.org/10.1103/PhysRevD.46.5346}{\emph{Phys. Rev. D}
  {\bfseries 46} (1992) 5346}
  [\href{https://arxiv.org/abs/hep-ph/9209238}{{\ttfamily hep-ph/9209238}}].

\bibitem{Barnaby:2010vf}
N.~Barnaby and M.~Peloso, \emph{{Large Nongaussianity in Axion Inflation}},
  \href{https://doi.org/10.1103/PhysRevLett.106.181301}{\emph{Phys. Rev. Lett.}
  {\bfseries 106} (2011) 181301}
  [\href{https://arxiv.org/abs/1011.1500}{{\ttfamily 1011.1500}}].

\bibitem{Adshead:2013qp}
P.~Adshead, E.~Martinec and M.~Wyman, \emph{{Gauge fields and inflation: Chiral
  gravitational waves, fluctuations, and the Lyth bound}},
  \href{https://doi.org/10.1103/PhysRevD.88.021302}{\emph{Phys. Rev. D}
  {\bfseries 88} (2013) 021302}
  [\href{https://arxiv.org/abs/1301.2598}{{\ttfamily 1301.2598}}].

\bibitem{Cheng:2015oqa}
S.-L.~Cheng, W.~Lee and K.-W.~Ng, \emph{{Numerical study of pseudoscalar
  inflation with an axion-gauge field coupling}},
  \href{https://doi.org/10.1103/PhysRevD.93.063510}{\emph{Phys. Rev. D}
  {\bfseries 93} (2016) 063510}
  [\href{https://arxiv.org/abs/1508.00251}{{\ttfamily 1508.00251}}].

\bibitem{Barnaby:2011vw}
N.~Barnaby, R.~Namba and M.~Peloso, \emph{{Phenomenology of a Pseudo-Scalar
  Inflaton: Naturally Large Nongaussianity}},
  \href{https://doi.org/10.1088/1475-7516/2011/04/009}{\emph{JCAP} {\bfseries
  04} (2011) 009} [\href{https://arxiv.org/abs/1102.4333}{{\ttfamily
  1102.4333}}].

\bibitem{Cook:2011hg}
J.L.~Cook and L.~Sorbo, \emph{{Particle production during inflation and
  gravitational waves detectable by ground-based interferometers}},
  \href{https://doi.org/10.1103/PhysRevD.85.023534}{\emph{Phys. Rev. D}
  {\bfseries 85} (2012) 023534}
  [\href{https://arxiv.org/abs/1109.0022}{{\ttfamily 1109.0022}}].

\bibitem{Barnaby:2011qe}
N.~Barnaby, E.~Pajer and M.~Peloso, \emph{{Gauge Field Production in Axion
  Inflation: Consequences for Monodromy, non-Gaussianity in the CMB, and
  Gravitational Waves at Interferometers}},
  \href{https://doi.org/10.1103/PhysRevD.85.023525}{\emph{Phys. Rev. D}
  {\bfseries 85} (2012) 023525}
  [\href{https://arxiv.org/abs/1110.3327}{{\ttfamily 1110.3327}}].

\bibitem{Sorbo:2011rz}
L.~Sorbo, \emph{{Parity violation in the Cosmic Microwave Background from a
  pseudoscalar inflaton}},
  \href{https://doi.org/10.1088/1475-7516/2011/06/003}{\emph{JCAP} {\bfseries
  06} (2011) 003} [\href{https://arxiv.org/abs/1101.1525}{{\ttfamily
  1101.1525}}].

\bibitem{Cook:2013xea}
J.L.~Cook and L.~Sorbo, \emph{{An inflationary model with small scalar and
  large tensor nongaussianities}},
  \href{https://doi.org/10.1088/1475-7516/2013/11/047}{\emph{JCAP} {\bfseries
  11} (2013) 047} [\href{https://arxiv.org/abs/1307.7077}{{\ttfamily
  1307.7077}}].

\bibitem{Bastero-Gil:2022fme}
M.~Bastero-Gil and A.T.~Manso, \emph{{Parity violating gravitational waves at
  the end of inflation}},
  \href{https://doi.org/10.1088/1475-7516/2023/08/001}{\emph{JCAP} {\bfseries
  08} (2023) 001} [\href{https://arxiv.org/abs/2209.15572}{{\ttfamily
  2209.15572}}].

\bibitem{Garcia-Bellido:2023ser}
J.~Garcia-Bellido, A.~Papageorgiou, M.~Peloso and L.~Sorbo, \emph{{A flashing
  beacon in axion inflation: recurring bursts of gravitational waves in the
  strong backreaction regime}},
  \href{https://doi.org/10.1088/1475-7516/2024/01/034}{\emph{JCAP} {\bfseries
  01} (2024) 034} [\href{https://arxiv.org/abs/2303.13425}{{\ttfamily
  2303.13425}}].

\bibitem{Adshead:2015pva}
P.~Adshead, J.T.~Giblin, T.R.~Scully and E.I.~Sfakianakis,
  \emph{{Gauge-preheating and the end of axion inflation}},
  \href{https://doi.org/10.1088/1475-7516/2015/12/034}{\emph{JCAP} {\bfseries
  12} (2015) 034} [\href{https://arxiv.org/abs/1502.06506}{{\ttfamily
  1502.06506}}].

\bibitem{Cuissa:2018oiw}
J.R.C.~Cuissa and D.G.~Figueroa, \emph{{Lattice formulation of axion inflation.
  Application to preheating}},
  \href{https://doi.org/10.1088/1475-7516/2019/06/002}{\emph{JCAP} {\bfseries
  06} (2019) 002} [\href{https://arxiv.org/abs/1812.03132}{{\ttfamily
  1812.03132}}].

\bibitem{Adshead:2023mvt}
P.~Adshead, J.T.~Giblin, R.~Grutkoski and Z.J.~Weiner, \emph{{Gauge preheating
  with full general relativity}},
  \href{https://doi.org/10.1088/1475-7516/2024/03/017}{\emph{JCAP} {\bfseries
  03} (2024) 017} [\href{https://arxiv.org/abs/2311.01504}{{\ttfamily
  2311.01504}}].

\bibitem{Adshead:2018doq}
P.~Adshead, J.T.~Giblin and Z.J.~Weiner, \emph{{Gravitational waves from gauge
  preheating}}, \href{https://doi.org/10.1103/PhysRevD.98.043525}{\emph{Phys.
  Rev. D} {\bfseries 98} (2018) 043525}
  [\href{https://arxiv.org/abs/1805.04550}{{\ttfamily 1805.04550}}].

\bibitem{Adshead:2019igv}
P.~Adshead, J.T.~Giblin, M.~Pieroni and Z.J.~Weiner, \emph{{Constraining Axion
  Inflation with Gravitational Waves across 29 Decades in Frequency}},
  \href{https://doi.org/10.1103/PhysRevLett.124.171301}{\emph{Phys. Rev. Lett.}
  {\bfseries 124} (2020) 171301}
  [\href{https://arxiv.org/abs/1909.12843}{{\ttfamily 1909.12843}}].

\bibitem{Adshead:2019lbr}
P.~Adshead, J.T.~Giblin, M.~Pieroni and Z.J.~Weiner, \emph{{Constraining axion
  inflation with gravitational waves from preheating}},
  \href{https://doi.org/10.1103/PhysRevD.101.083534}{\emph{Phys. Rev. D}
  {\bfseries 101} (2020) 083534}
  [\href{https://arxiv.org/abs/1909.12842}{{\ttfamily 1909.12842}}].

\bibitem{Notari:2016npn}
A.~Notari and K.~Tywoniuk, \emph{{Dissipative Axial Inflation}},
  \href{https://doi.org/10.1088/1475-7516/2016/12/038}{\emph{JCAP} {\bfseries
  12} (2016) 038} [\href{https://arxiv.org/abs/1608.06223}{{\ttfamily
  1608.06223}}].

\bibitem{DallAgata:2019yrr}
G.~Dall'Agata, S.~Gonz\'alez-Mart\'\i{}n, A.~Papageorgiou and M.~Peloso,
  \emph{{Warm dark energy}},
  \href{https://doi.org/10.1088/1475-7516/2020/08/032}{\emph{JCAP} {\bfseries
  08} (2020) 032} [\href{https://arxiv.org/abs/1912.09950}{{\ttfamily
  1912.09950}}].

\bibitem{Sobol:2019xls}
O.O.~Sobol, E.V.~Gorbar and S.I.~Vilchinskii, \emph{{Backreaction of
  electromagnetic fields and the Schwinger effect in pseudoscalar inflation
  magnetogenesis}},
  \href{https://doi.org/10.1103/PhysRevD.100.063523}{\emph{Phys. Rev. D}
  {\bfseries 100} (2019) 063523}
  [\href{https://arxiv.org/abs/1907.10443}{{\ttfamily 1907.10443}}].

\bibitem{Domcke:2020zez}
V.~Domcke, V.~Guidetti, Y.~Welling and A.~Westphal, \emph{{Resonant
  backreaction in axion inflation}},
  \href{https://doi.org/10.1088/1475-7516/2020/09/009}{\emph{JCAP} {\bfseries
  09} (2020) 009} [\href{https://arxiv.org/abs/2002.02952}{{\ttfamily
  2002.02952}}].

\bibitem{Gorbar:2021rlt}
E.V.~Gorbar, K.~Schmitz, O.O.~Sobol and S.I.~Vilchinskii, \emph{{Gauge-field
  production during axion inflation in the gradient expansion formalism}},
  \href{https://doi.org/10.1103/PhysRevD.104.123504}{\emph{Phys. Rev. D}
  {\bfseries 104} (2021) 123504}
  [\href{https://arxiv.org/abs/2109.01651}{{\ttfamily 2109.01651}}].

\bibitem{Peloso:2022ovc}
M.~Peloso and L.~Sorbo, \emph{{Instability in axion inflation with strong
  backreaction from gauge modes}},
  \href{https://doi.org/10.1088/1475-7516/2023/01/038}{\emph{JCAP} {\bfseries
  01} (2023) 038} [\href{https://arxiv.org/abs/2209.08131}{{\ttfamily
  2209.08131}}].

\bibitem{Durrer:2023rhc}
R.~Durrer, O.~Sobol and S.~Vilchinskii, \emph{{Backreaction from gauge fields
  produced during inflation}},
  \href{https://doi.org/10.1103/PhysRevD.108.043540}{\emph{Phys. Rev. D}
  {\bfseries 108} (2023) 043540}
  [\href{https://arxiv.org/abs/2303.04583}{{\ttfamily 2303.04583}}].

\bibitem{vonEckardstein:2023gwk}
R.~von Eckardstein, M.~Peloso, K.~Schmitz, O.~Sobol and L.~Sorbo, \emph{{Axion
  inflation in the strong-backreaction regime: decay of the Anber-Sorbo
  solution}}, \href{https://doi.org/10.1007/JHEP11(2023)183}{\emph{JHEP}
  {\bfseries 11} (2023) 183}
  [\href{https://arxiv.org/abs/2309.04254}{{\ttfamily 2309.04254}}].

\bibitem{Galanti:2024jhw}
D.C.~Galanti, P.~Conzinu, G.~Marozzi and S.~Santos~da Costa, \emph{{Gauge
  invariant quantum backreaction in U(1) axion inflation}},
  \href{https://doi.org/10.1103/PhysRevD.110.123510}{\emph{Phys. Rev. D}
  {\bfseries 110} (2024) 123510}
  [\href{https://arxiv.org/abs/2406.19960}{{\ttfamily 2406.19960}}].

\bibitem{Durrer:2024ibi}
R.~Durrer, R.~von Eckardstein, D.~Garg, K.~Schmitz, O.~Sobol and
  S.~Vilchinskii, \emph{{Scalar perturbations from inflation in the presence of
  gauge fields}},
  \href{https://doi.org/10.1103/PhysRevD.110.043533}{\emph{Phys. Rev. D}
  {\bfseries 110} (2024) 043533}
  [\href{https://arxiv.org/abs/2404.19694}{{\ttfamily 2404.19694}}].

\bibitem{vonEckardstein:2024tix}
R.~von Eckardstein, K.~Schmitz and O.~Sobol, \emph{{On the Schwinger effect
  during axion inflation}},
  \href{https://doi.org/10.1007/JHEP02(2025)096}{\emph{JHEP} {\bfseries 02}
  (2025) 096} [\href{https://arxiv.org/abs/2408.16538}{{\ttfamily
  2408.16538}}].

\bibitem{Caravano:2021bfn}
A.~Caravano, E.~Komatsu, K.D.~Lozanov and J.~Weller, \emph{{Lattice simulations
  of Abelian gauge fields coupled to axions during inflation}},
  \href{https://doi.org/10.1103/PhysRevD.105.123530}{\emph{Phys. Rev. D}
  {\bfseries 105} (2022) 123530}
  [\href{https://arxiv.org/abs/2110.10695}{{\ttfamily 2110.10695}}].

\bibitem{Caravano:2022epk}
A.~Caravano, E.~Komatsu, K.D.~Lozanov and J.~Weller, \emph{{Lattice simulations
  of axion-U(1) inflation}},
  \href{https://doi.org/10.1103/PhysRevD.108.043504}{\emph{Phys. Rev. D}
  {\bfseries 108} (2023) 043504}
  [\href{https://arxiv.org/abs/2204.12874}{{\ttfamily 2204.12874}}].

\bibitem{Figueroa:2023oxc}
D.G.~Figueroa, J.~Lizarraga, A.~Urio and J.~Urrestilla, \emph{{Strong
  Backreaction Regime in Axion Inflation}},
  \href{https://doi.org/10.1103/PhysRevLett.131.151003}{\emph{Phys. Rev. Lett.}
  {\bfseries 131} (2023) 151003}
  [\href{https://arxiv.org/abs/2303.17436}{{\ttfamily 2303.17436}}].

\bibitem{Figueroa:2024rkr}
D.G.~Figueroa, J.~Lizarraga, N.~Loayza, A.~Urio and J.~Urrestilla,
  \emph{{Nonlinear dynamics of axion inflation: A detailed lattice study}},
  \href{https://doi.org/10.1103/PhysRevD.111.063545}{\emph{Phys. Rev. D}
  {\bfseries 111} (2025) 063545}
  [\href{https://arxiv.org/abs/2411.16368}{{\ttfamily 2411.16368}}].

\bibitem{Sharma:2024nfu}
R.~Sharma, A.~Brandenburg, K.~Subramanian and A.~Vikman, \emph{{Lattice
  simulations of axion-U(1) inflation: gravitational waves, magnetic fields,
  and scalar statistics}},  \href{https://arxiv.org/abs/2411.04854}{{\ttfamily
  2411.04854}}.

\bibitem{BICEP:2021xfz}
{\scshape BICEP, Keck} collaboration, \emph{{Improved Constraints on Primordial
  Gravitational Waves using Planck, WMAP, and BICEP/Keck Observations through
  the 2018 Observing Season}},
  \href{https://doi.org/10.1103/PhysRevLett.127.151301}{\emph{Phys. Rev. Lett.}
  {\bfseries 127} (2021) 151301}
  [\href{https://arxiv.org/abs/2110.00483}{{\ttfamily 2110.00483}}].

\bibitem{Meerburg:2012id}
P.D.~Meerburg and E.~Pajer, \emph{{Observational Constraints on Gauge Field
  Production in Axion Inflation}},
  \href{https://doi.org/10.1088/1475-7516/2013/02/017}{\emph{JCAP} {\bfseries
  02} (2013) 017} [\href{https://arxiv.org/abs/1203.6076}{{\ttfamily
  1203.6076}}].

\bibitem{Linde:2012bt}
A.~Linde, S.~Mooij and E.~Pajer, \emph{{Gauge field production in supergravity
  inflation: Local non-Gaussianity and primordial black holes}},
  \href{https://doi.org/10.1103/PhysRevD.87.103506}{\emph{Phys. Rev. D}
  {\bfseries 87} (2013) 103506}
  [\href{https://arxiv.org/abs/1212.1693}{{\ttfamily 1212.1693}}].

\bibitem{Anber:2015yca}
M.M.~Anber and E.~Sabancilar, \emph{{Hypermagnetic Fields and Baryon Asymmetry
  from Pseudoscalar Inflation}},
  \href{https://doi.org/10.1103/PhysRevD.92.101501}{\emph{Phys. Rev. D}
  {\bfseries 92} (2015) 101501}
  [\href{https://arxiv.org/abs/1507.00744}{{\ttfamily 1507.00744}}].

\bibitem{Domcke:2018eki}
V.~Domcke and K.~Mukaida, \emph{{Gauge Field and Fermion Production during
  Axion Inflation}},
  \href{https://doi.org/10.1088/1475-7516/2018/11/020}{\emph{JCAP} {\bfseries
  11} (2018) 020} [\href{https://arxiv.org/abs/1806.08769}{{\ttfamily
  1806.08769}}].

\bibitem{Domcke:2019qmm}
V.~Domcke, Y.~Ema and K.~Mukaida, \emph{{Chiral Anomaly, Schwinger Effect,
  Euler-Heisenberg Lagrangian, and application to axion inflation}},
  \href{https://doi.org/10.1007/JHEP02(2020)055}{\emph{JHEP} {\bfseries 02}
  (2020) 055} [\href{https://arxiv.org/abs/1910.01205}{{\ttfamily
  1910.01205}}].

\bibitem{Cado:2022pxk}
Y.~Cado and M.~Quir\'os, \emph{{Numerical study of the Schwinger effect in
  axion inflation}},
  \href{https://doi.org/10.1103/PhysRevD.106.123527}{\emph{Phys. Rev. D}
  {\bfseries 106} (2022) 123527}
  [\href{https://arxiv.org/abs/2208.10977}{{\ttfamily 2208.10977}}].

\bibitem{Domcke:2023tnn}
V.~Domcke, Y.~Ema and S.~Sandner, \emph{{Perturbatively including
  inhomogeneities in axion inflation}},
  \href{https://doi.org/10.1088/1475-7516/2024/03/019}{\emph{JCAP} {\bfseries
  03} (2024) 019} [\href{https://arxiv.org/abs/2310.09186}{{\ttfamily
  2310.09186}}].

\bibitem{Fujita:2025zoa}
T.~Fujita, K.~Mukaida and T.~Tsuji, \emph{{Reheating after Axion Inflation}},
  \href{https://arxiv.org/abs/2503.01228}{{\ttfamily 2503.01228}}.

\bibitem{He:2025ieo}
J.-F.~He, K.-G.~Zhang, C.~Fu and Z.-K.~Guo, \emph{{Strong backreaction of gauge
  quanta produced during inflation}},
  \href{https://doi.org/10.1103/PhysRevD.111.103525}{\emph{Phys. Rev. D}
  {\bfseries 111} (2025) 103525}
  [\href{https://arxiv.org/abs/2502.13158}{{\ttfamily 2502.13158}}].

\bibitem{Kume:2025lvz}
J.~Kume, M.~Peloso and N.~Bartolo, \emph{{Revisiting the Chern-Simons
  interaction during inflation with a non-canonical pseudo-scalar}},
  \href{https://arxiv.org/abs/2501.02890}{{\ttfamily 2501.02890}}.

\bibitem{Corba:2025reo}
S.P.~Corb\`a, \emph{{Gravitational wave anisotropies from axion inflation}},
  \href{https://arxiv.org/abs/2504.13156}{{\ttfamily 2504.13156}}.

\bibitem{Ozsoy:2024apn}
O.~\"Ozsoy, A.~Papageorgiou and M.~Fasiello, \emph{{Scale-dependent chirality
  as a smoking gun for Abelian gauge fields during inflation}},
  \href{https://doi.org/10.1088/1475-7516/2024/12/008}{\emph{JCAP} {\bfseries
  12} (2024) 008} [\href{https://arxiv.org/abs/2405.14963}{{\ttfamily
  2405.14963}}].

\bibitem{Corba:2024tfz}
S.P.~Corb\`a and L.~Sorbo, \emph{{Correlated scalar perturbations and
  gravitational waves from axion inflation}},
  \href{https://doi.org/10.1088/1475-7516/2024/10/024}{\emph{JCAP} {\bfseries
  10} (2024) 024} [\href{https://arxiv.org/abs/2403.03338}{{\ttfamily
  2403.03338}}].

\bibitem{Gorbar:2023zla}
E.V.~Gorbar, A.I.~Momot, O.O.~Prikhodko and O.M.~Teslyk, \emph{{Hydrodynamical
  approach to chirality production during axion inflation}},
  \href{https://doi.org/10.1103/PhysRevD.109.023536}{\emph{Phys. Rev. D}
  {\bfseries 109} (2024) 023536}
  [\href{https://arxiv.org/abs/2311.07429}{{\ttfamily 2311.07429}}].

\bibitem{Unal:2023srk}
C.~Unal, A.~Papageorgiou and I.~Obata, \emph{{Axion-gauge dynamics during
  inflation as the origin of pulsar timing array signals and primordial black
  holes}}, \href{https://doi.org/10.1016/j.physletb.2024.138873}{\emph{Phys.
  Lett. B} {\bfseries 856} (2024) 138873}
  [\href{https://arxiv.org/abs/2307.02322}{{\ttfamily 2307.02322}}].

\bibitem{Figueroa:2017qmv}
D.G.~Figueroa and M.~Shaposhnikov, \emph{{Lattice implementation of Abelian
  gauge theories with Chern\textendash{}Simons number and an axion field}},
  \href{https://doi.org/10.1016/j.nuclphysb.2017.12.001}{\emph{Nucl. Phys. B}
  {\bfseries 926} (2018) 544}
  [\href{https://arxiv.org/abs/1705.09629}{{\ttfamily 1705.09629}}].

\bibitem{Carpenter1994Thirdorder2R}
M.H.~Carpenter and C.A.~Kennedy, \emph{Third-order 2n-storage runge-kutta
  schemes with error control},  1994,
  \href{https://api.semanticscholar.org/CorpusID:118434708}{https://api.semanticscholar.org/CorpusID:118434708}.

\bibitem{Carpenter1994Fourthorder2R}
M.H.~Carpenter and C.A.~Kennedy, \emph{Fourth-order 2n-storage runge-kutta
  schemes},  1994,
  \href{https://api.semanticscholar.org/CorpusID:116658826}{https://api.semanticscholar.org/CorpusID:116658826}.

\bibitem{Figueroa:2021iwm}
D.G.~Figueroa, A.~Florio, T.~Opferkuch and B.A.~Stefanek, \emph{{Lattice
  simulations of non-minimally coupled scalar fields in the Jordan frame}},
  \href{https://doi.org/10.21468/SciPostPhys.15.3.077}{\emph{SciPost Phys.}
  {\bfseries 15} (2023) 077}
  [\href{https://arxiv.org/abs/2112.08388}{{\ttfamily 2112.08388}}].

\bibitem{Figueroa:2020rrl}
D.G.~Figueroa, A.~Florio, F.~Torrenti and W.~Valkenburg, \emph{{The art of
  simulating the early Universe -- Part I}},
  \href{https://doi.org/10.1088/1475-7516/2021/04/035}{\emph{JCAP} {\bfseries
  04} (2021) 035} [\href{https://arxiv.org/abs/2006.15122}{{\ttfamily
  2006.15122}}].

\bibitem{Figueroa:2021yhd}
D.G.~Figueroa, A.~Florio, F.~Torrenti and W.~Valkenburg, \emph{{CosmoLattice: A
  modern code for lattice simulations of scalar and gauge field dynamics in an
  expanding universe}},
  \href{https://doi.org/10.1016/j.cpc.2022.108586}{\emph{Comput. Phys. Commun.}
  {\bfseries 283} (2023) 108586}
  [\href{https://arxiv.org/abs/2102.01031}{{\ttfamily 2102.01031}}].

\bibitem{ACT:2025tim}
{\scshape ACT} collaboration, \emph{{The Atacama Cosmology Telescope: DR6
  Constraints on Extended Cosmological Models}},
  \href{https://arxiv.org/abs/2503.14454}{{\ttfamily 2503.14454}}.

\bibitem{ACT:2025fju}
{\scshape ACT} collaboration, \emph{{The Atacama Cosmology Telescope: DR6 Power
  Spectra, Likelihoods and $\Lambda$CDM Parameters}},
  \href{https://arxiv.org/abs/2503.14452}{{\ttfamily 2503.14452}}.

\bibitem{Figueroa:2024yja}
D.G.~Figueroa and N.~Loayza, \emph{{Geometric reheating of the Universe}},
  \href{https://doi.org/10.1088/1475-7516/2025/03/073}{\emph{JCAP} {\bfseries
  03} (2025) 073} [\href{https://arxiv.org/abs/2406.02689}{{\ttfamily
  2406.02689}}].

\bibitem{Linde:1983gd}
A.D.~Linde, \emph{{Chaotic Inflation}},
  \href{https://doi.org/10.1016/0370-2693(83)90837-7}{\emph{Phys. Lett. B}
  {\bfseries 129} (1983) 177}.

\bibitem{Freese:1993bc}
K.~Freese, \emph{{Natural Inflation}},  in \emph{{37th Yamada Conference:
  Evolution of the Universe and its Observational Quest}}, pp.~49--58, 6, 1993
  [\href{https://arxiv.org/abs/astro-ph/9310012}{{\ttfamily
  astro-ph/9310012}}].

\bibitem{Silverstein:2008sg}
E.~Silverstein and A.~Westphal, \emph{{Monodromy in the CMB: Gravity Waves and
  String Inflation}},
  \href{https://doi.org/10.1103/PhysRevD.78.106003}{\emph{Phys. Rev. D}
  {\bfseries 78} (2008) 106003}
  [\href{https://arxiv.org/abs/0803.3085}{{\ttfamily 0803.3085}}].

\bibitem{Starobinsky:1980te}
A.A.~Starobinsky, \emph{{A New Type of Isotropic Cosmological Models Without
  Singularity}},
  \href{https://doi.org/10.1016/0370-2693(80)90670-X}{\emph{Phys. Lett. B}
  {\bfseries 91} (1980) 99}.

\bibitem{Maeda:1988ab}
K.-i.~Maeda, \emph{{Towards the Einstein-Hilbert Action via Conformal
  Transformation}}, \href{https://doi.org/10.1103/PhysRevD.39.3159}{\emph{Phys.
  Rev. D} {\bfseries 39} (1989) 3159}.

\bibitem{Whitt:1984pd}
B.~Whitt, \emph{{Fourth Order Gravity as General Relativity Plus Matter}},
  \href{https://doi.org/10.1016/0370-2693(84)90332-0}{\emph{Phys. Lett. B}
  {\bfseries 145} (1984) 176}.

\bibitem{Kallosh:2013hoa}
R.~Kallosh and A.~Linde, \emph{{Universality Class in Conformal Inflation}},
  \href{https://doi.org/10.1088/1475-7516/2013/07/002}{\emph{JCAP} {\bfseries
  07} (2013) 002} [\href{https://arxiv.org/abs/1306.5220}{{\ttfamily
  1306.5220}}].

\bibitem{Kallosh:2013yoa}
R.~Kallosh, A.~Linde and D.~Roest, \emph{{Superconformal Inflationary
  $\alpha$-Attractors}},
  \href{https://doi.org/10.1007/JHEP11(2013)198}{\emph{JHEP} {\bfseries 11}
  (2013) 198} [\href{https://arxiv.org/abs/1311.0472}{{\ttfamily 1311.0472}}].

\bibitem{Galante:2014ifa}
M.~Galante, R.~Kallosh, A.~Linde and D.~Roest, \emph{{Unity of Cosmological
  Inflation Attractors}},
  \href{https://doi.org/10.1103/PhysRevLett.114.141302}{\emph{Phys. Rev. Lett.}
  {\bfseries 114} (2015) 141302}
  [\href{https://arxiv.org/abs/1412.3797}{{\ttfamily 1412.3797}}].

\bibitem{Kallosh:2015lwa}
R.~Kallosh and A.~Linde, \emph{{Planck, LHC, and $\alpha$-attractors}},
  \href{https://doi.org/10.1103/PhysRevD.91.083528}{\emph{Phys. Rev. D}
  {\bfseries 91} (2015) 083528}
  [\href{https://arxiv.org/abs/1502.07733}{{\ttfamily 1502.07733}}].

\bibitem{Boubekeur:2005zm}
L.~Boubekeur and D.H.~Lyth, \emph{{Hilltop inflation}},
  \href{https://doi.org/10.1088/1475-7516/2005/07/010}{\emph{JCAP} {\bfseries
  07} (2005) 010} [\href{https://arxiv.org/abs/hep-ph/0502047}{{\ttfamily
  hep-ph/0502047}}].

\bibitem{Kallosh:2019jnl}
R.~Kallosh and A.~Linde, \emph{{On hilltop and brane inflation after Planck}},
  \href{https://doi.org/10.1088/1475-7516/2019/09/030}{\emph{JCAP} {\bfseries
  09} (2019) 030} [\href{https://arxiv.org/abs/1906.02156}{{\ttfamily
  1906.02156}}].

\end{thebibliography}\endgroup
\end{document}